\documentclass{aa}  

\usepackage{graphicx}

\usepackage{txfonts}

\usepackage{pdflscape}
\usepackage{comment}
\usepackage{orcidlink}
\usepackage{multirow}
\usepackage[switch]{lineno}
\usepackage{hyperref}

\newcommand{\Halpha}{\ion{H}{$\alpha$}}
\newcommand{\Hbeta}{\ion{H}{$\beta$}}

\newcommand{\HeI}{\ion{He}{I}}

\newcommand{\ztfg}{$g$}
\newcommand{\ztfr}{$r$}
\newcommand{\ztfi}{$i$}

\begin{document} 

   \title{ZTF-observed late-time signals of pre-ZTF transients}

   \author{Jacco H. Terwel \orcidlink{0000-0001-9834-3439} \inst{1,2}
   \and Kate Maguire \orcidlink{0000-0002-9770-3508} \inst{1} 
   \and Jesper Sollerman \orcidlink{0000-0003-1546-6615} \inst{3}
   \and Phil Wiseman \orcidlink{0000-0002-3073-1512} \inst{4}
   \and Tracy X. Chen \orcidlink{0000-0001-9152-6224} \inst{5}
   \and Matthew J. Graham \orcidlink{0000-0002-3168-0139} \inst{6}
   \and Russ R. Laher \orcidlink{0000-0003-2451-5482} \inst{5}
   \and Reed Riddle \orcidlink{0000-0002-0387-370X} \inst{6}
   \and Niharika Sravan \inst{7}
          }

   \institute{School of Physics, Trinity College Dublin, The University of Dublin, Dublin 2, Ireland\\
   \email{terwelj@tcd.ie}
   \and Nordic Optical Telescope, Rambla José Ana Fernández Pérez 7, ES-38711 Breña Baja, Spain
   \and Oskar Klein Centre, Department of Astronomy, Stockholm University, Albanova University Center, SE-106 91, Stockholm, Sweden
   \and School of Physics and Astronomy, University of Southampton,  Southampton, SO17 1BJ, UK
   \and IPAC, California Institute of Technology, 1200 E. California Blvd, Pasadena, CA, 91125, USA
   \and Department Physics, Math, and Astronomy, California Institute of Technology, 1200 E. California Blvd, Pasadena, CA, 91125, USA
   \and Department of Physics, Drexel University, Philadelphia, PA 19104, USA
             }

   \date{Received XXX; accepted YYY}

  \abstract 
   {With large-scale surveys such as the Zwicky Transient Facility (ZTF), it has become possible to obtain a well-sampled light curve spanning the full length of the survey for any discovery within the survey footprint. Similarly, any transient within the footprint that was first detected before the start of the survey will likely have a large number of post-transient observations, making them excellent targets to search for the presence of late-time signals, in particular due to interaction with circumstellar material (CSM). We search for late-time signals in a sample of 7\,718 transients, mainly supernovae (SNe), which were first detected during the 10 years before the start of ZTF, aiming to find objects showing signs of late-time interaction with CSM. We find one candidate whose late-time signal is best explained by late-time CSM interaction, with the signal being around 300 days after transient discovery. A thin, distant shell containing $\lesssim5$ M$_\odot$ of material could explain the recovered signal. We also find five objects whose late-time signal is best explained by faint nuclear transients occurring in host nuclei close to the pre-ZTF transient locations. Finally, we find two objects where it is difficult to determine whether the signal is from a nuclear transient or due to late-time CSM interaction $>5$ years after the SN. This study demonstrates the ability of large-scale surveys to find faint transient signals for a variety of objects, uncovering a population of previously unknown sources. However, the large number of non-detections show that strong late-time CSM interaction occurring years after the SN explosion is extremely rare.}


   \maketitle

\section{Introduction}
Transients are, by definition, events that last for a limited amount of time. A well-known class of transients are supernovae (SNe): terminal explosions of stars at the end of their lives that are bright enough to outshine their host galaxies. After the explosion the SN rises rapidly to its peak brightness before fading away over the next few weeks to months, sometimes showing a plateau before fading away further. SNe arise from the explosions of massive stars or thermonuclear explosions of white dwarfs (WDs) but there are many variations \citep[e.g.~][]{Galyam_HSN, Jha_2019_review, Liu_Iareview}. Other types of extragalactic transients exist besides SNe, some of which have a strong preference for galactic nuclei. 
In the effort to systematically discover, observe, and follow up on transient events, dedicated large transient surveys such as the Panoramic Survey Telescope and Rapid Response System (Pan-STARRS, \citealt{Pan-STARRS1}, 2008 -- present), (intermediate) Palomar Transient Factory (PTF, \citealt{PTF_1, PTF_2}, 2009 -- 2012, iPTF, \citealt{iPTF}, 2013 -- 2016), All Sky Automated Survey for SuperNovae (ASASSN, \citealt{ASASSN_paper1, ASASSN_catalog}, 2013 -- present), Asteroid Terrestrial-impact Last Alert System (ATLAS, \citealt{ATLAS}, 2015 - present), and Zwicky Transient Facility (ZTF, \citealt{ZTF_Surveys_Scheduler, ZTF_overview_and_1st_results, ZTF_Science_Objectives, ZTF_Instrumentation, ZTF_Observing_System}, 2018 -- present) regularly observe the same part of the sky to register any changing object. Other sky surveys such as Gaia (\citealt{Gaia}, 2014 -- present) may also observe transients that happened to be in their field of view, even though their main mission is not oriented towards transients.

Most transient research focuses on young transients up to and around peak brightness, as these are the phases where they evolve the fastest and show features that are only visible for a short time after the explosion. At the earliest phases, only the outermost ejecta of SNe can be observed, but as they expand and cool the inner, slower ejecta are revealed \citep{Jerkstrand_2017HSN_nebular}. This allows us to get a radial abundance profile of the ejected elements, giving crucial information on the composition of the progenitor. The decline of the light curve can be slowed or temporarily stalled by interaction between the SN ejecta and circumstellar material \citep[CSM;][]{Blinnikov_2017HSN_interactingSNe, Chevalier_2017HSN_CSM}, resulting in very long-lived transients.

The most commonly discovered type of SNe are SNe Ia, which are well known as standardisable candles as their peak absolute magnitudes can be standardised based on their light curve properties to measure the distances to them \citep[e.g.][]{Phillips_rel, Phillips_rel2}. However, it is still not clear what the progenitor systems of these explosions are. In the double degenerate scenario, two WDs are expected to interact or merge before exploding \citep{Iben_Double_degenerate, Webbink_Double_degenerate, Pakmor_merger, Shen_D6}. If the donor star is not degenerate, it can donate hydrogen-rich material through mass transfer until the WD explodes \citep{Whelan_classical_Ia_mod, Nomoto_single_degenerate}. Besides normal SNe Ia, there are many subclasses that cannot be standardised due to photometric and/or spectroscopic differences. 

One subclass of SNe Ia (`Ia-CSM') begins to show emission lines around or shortly after their peak that are usually not present in SNe Ia, with the most prominent of these being \Halpha\ and \Hbeta\ (e.g., SN 2002ic \citealt{02ic_H_det, Hamuy_02ic}, PTF11kx \citealt{PTF11kx}), or \HeI\ in the case of SN 2020eyj \citep{Kool_He_CSM}. These signatures are thought to originate from interaction between the SN ejecta and pre-existing CSM. Alongside these spectral signatures, the light curves of these objects tend to decline extremely slowly or even plateau for up to several hundreds of days \citep[e.g.][]{02ic_slow_decay, Ia-CSM_BTS}. Ia-CSM are very rare ($\sim$3\% of SNe Ia, \citealt{DR2_diversity}) and to this day only a few dozen examples are known \citep{2005gj, Ia-CSM_Silverman, Ia-CSM_BTS}, some of which are contested with a SN IIn classification due to their similar signatures. Some double-degenerate models are able to produce a limited amount of CSM \citep{Double_degen_CSM_gen}, but this is generally not enough to generate a detectable signal. Previous estimates of CSM shell masses for known or suspected SNe Ia-CSM events have been up to $\sim 5 M_\odot$ \citep{Chugai_2004, PTF11kx, Inserra_2016}. Besides this, the presence of \Halpha~in their spectra suggests a single degenerate progenitor system.

Signatures of delayed CSM interaction in SNe Ia have also been studied by systematically targeting $\geq 1$ year old SNe using the \textit{Hubble Space Telescope (HST)} \citep{2015cp}, resulting in the discovery of SN 2015cp showing signs of CSM interaction over 1.5 years after explosion. \citet{GALEX_Late_CSM} looked at archival data from the \textit{Galactic Evolution Explorer (GALEX)} for a sample of 1080 SNe Ia, finding no signs of late-time CSM interaction in any of them. From this, they estimated that late-time CSM interaction with the strength of SN 2015cp occurs in $<5$ per cent of SNe Ia. \citet{Terwel_2024_paper1} looked for signatures of late-time ($>$100 d after peak) rebrightening in 3\,628 SNe Ia discovered by ZTF between 2018 and 2020. They identified three objects with late-time rebrightening between 500 and 1500 days after the peak that cannot be easily explained by other means (such as an unresolved sibling transient, host activity, or data issues). They derived a rate of strong late-time CSM interaction of $8_{-4}^{+20}$ to $54_{-26}^{+91}$ Gpc$^{-3}$ yr$^{-1}$, assuming a constant SN Ia for $z \leq\ 0.1$ \citep{Frohmaier2019}.

Several classes of core-collapse SNe also show signatures of interaction, see \citet{Ibn_IIn_SNe_handbook} for an overview. SNe Ibn and IIn are classes of events characterised by strong interaction with a dense CSM ejected within the last years before the explosion. SNe II can also show narrow emission line features that disappear within hours to days after the explosion (e.g. \citealt{Bruch_early_lines_SNII, 2023ixf_Erez}), which suggests interaction with CSM that was lost shortly before the SN occurred \citep{Confined_SNII_CSM}. Some superluminous SNe (SLSNe) are thought to arise from CSM interaction as well \citep{Late-time_CSM_SLSNE_I}, with the extra energy source pushing the brightness significantly above that of their normal counterparts. All of these types of SNe, as well as those that are not known to interact at early phases, could have distant shells of CSM causing a long delay between the explosion and the start of the CSM interaction.

There are other types of objects that can appear as transients and have signatures that could last for many years. Active galactic nuclei (AGN) are also known to vary with time. In some cases AGN variability is so extreme that it changes the entire spectrum of the AGN, with (dis)appearing broad emission lines and continuum flux. These are so-called changing-look AGN (CL-AGN) \citep[see][for a review]{CLAGN}. Some nuclear variability does not quite fit the known classes of AGN \citep{Antonucci_1993_AGN, Urry_1995_AGN} or tidal disruption events \citep[TDEs;][]{Rees_1988_TDE, Strubbe_2009_TDE}. These events have been named `ambiguous nuclear transients' \citep[ANTS;][]{Kankare_ANT, 2020ohl_Hinkle, Hinkle_MIR_ANT_echo, Hinkle_Extreme_nuclear_transients/ANTs, wiseman_ztfants} and rise quickly within a few weeks before declining very slowly over hundreds of days (although their decline rates are variable). 

With a covered area of 25\,000 to 30\,000 square degrees (the entire northern sky above dec $\sim -30^{\circ}$), observing in three broadband optical filters (ZTF-\ztfg, ZTF-\ztfr, ZTF-\ztfi) with a two to three day cadence, and relatively deep limits of $\sim20.5$ mag, ZTF is ideal for finding many transients and allows us to study them over their full evolution, do statistics on samples, and find rare subclasses of events. Every object has a light curve spanning the full operation time of the survey, including pre-discovery observations which can be used to search progenitor activity \citep{PTF_IIb_Precursor_search, stacked_precursors, ZTF_SN_progenitor_search} and observations long after the SN to monitor its long-term evolution.

In this paper, we search for ZTF-detected late-time signals of transients that were first detected between 1 January 2008 and 31 December 2017. ZTF has an excellent catalogue of late-time observations of these pre-ZTF transients spanning six years. Using a similar methodology to \citet{Terwel_2024_paper1}, we bin the ZTF observations in bins of 25 to 100 days to push the detection limit of the late-time observations up to  1 mag beyond that of the individual exposures. Any potential candidate with a late-time signal is then further investigated. In Section~\ref{data}, we build our sample of objects whose first detection was in the decade before the start of ZTF. In Section~\ref{analysis}, we introduce the changes we made to the pipeline from \citet{Terwel_2024_paper1} to adapt it to our sample, and the post-binning analysis. In Section~\ref{results}, we describe the objects flagged by the binning program and split them into different groups with different origins. These are discussed in Section~\ref{discussion}, and we conclude in Section~\ref{conclusions}. Throughout this paper, to convert between apparent and absolute magnitude we assume a flat $\Lambda$CDM cosmology with a Hubble Constant, H$_0 = 67.7$ km s$^{-1}$ Mpc$^{-1}$, and $\Omega_\text{m} = 0.310$ \citep{Planck18VI}.

\section{Data}
\label{data}
Our first aim is to compile a list of all transients discovered in the decade before ZTF started. To build our sample, we started with all transients in the Open Supernova Catalog (OSC\footnote{\url{https://github.com/astrocatalogs/supernovae}}, \citealt{Open_SN_cat}) that were discovered between 1 January 2008 up to 1 January 2018, giving us 22\,790 objects. We end our sample a few months before the start of the full ZTF survey in March 2018 to remove most pre-ZTF transients that were still visible by the start of ZTF. For each object we require a name, sky position, redshift, and classification of its type. We require the redshift to make an informed estimate of the absolute luminosity of any potentially detected late-time signal. For the vast majority of transients, a classification means a spectrum was obtained that enabled typing, but the size of the sample precludes checking of each individual classification. However, further investigation of identified interesting events is performed where required. 

The OSC includes `SN candidate' as a possible transient type. These are likely SNe that were never spectroscopically classified, although this group of objects is likely to be contaminated with some non-SN transients. We include this group of objects without spectra, despite lacking a more definite classification, as this group is relatively small and we want to be as inclusive as possible. These requirements cut our sample down to 8\,865 objects. By querying WISeREP\footnote{\url{https://www.wiserep.org}} \citep{wiserep} for objects reported in the same date range, we attempted to recover objects that were incomplete or absent from the OSC. Out of the 12\,955 OSC objects that did not have all the required information and no WISeREP match, 1\,137 had no reported sky position, while the other 11\,818 had a sky position but no reported redshift and/or classification.

From the combined sample, we removed objects at Dec $\leq -32\deg$ as ZTF is unable to observe below Dec $\approx -30\deg$. After these updates (including WISeREP and with the declination cut), our sample increased to 8\,914 objects. The declination cut is liberal to ensure all object locations that could have been observed with ZTF remain in our sample. Objects outside the ZTF footprint but still in our sample are cut automatically when obtaining the ZTF light curves. We do not make a cut on proximity to the host nucleus as we want to be as inclusive as possible. Instead, we accept the possible contamination by nuclear activity which, as will be shown later, is in most cases identifiable.

While cross-matching between OSC and WISeREP to remove duplicate entries, we noted several objects with different names and discovery dates but located very close on the sky. While uncommon, it is possible for multiple SNe to occur at the same sky position in short succession. \citet{Terwel_2024_paper1} found several such sibling transients at very small spatial separations, where both SNe were detected by ZTF. It could also mean that the two transients are connected, such as a precursor event and a true SN explosion \citep[see e.g.][for a systematic search for this type of event]{stacked_precursors}. Since we just look at the late-time light curve at the sky position of each pre-ZTF transient, sibling transients like these are investigated together as they have the same sky position.

Our late-time light curves at the position of each transient in the list were constructed, using the method detailed in Section \ref{analysis}, between 9 December 2023 and 24 January 2024 using all available ZTF data at the time. This results in an effective range of over 5.5 years of ZTF observations for each object in our sample. For 207 sky positions that lay close to the edge of the ZTF survey, no ZTF observations were obtained and these were removed from our sample, leaving us with 8\,707 transients. The ZTF light curves at these sky positions are generated using \textsc{fpbot}\footnote{\url{https://github.com/simeonreusch/fpbot}} \citep{fpbot}.

\begin{figure}
    \centering
    \includegraphics[width=\columnwidth]{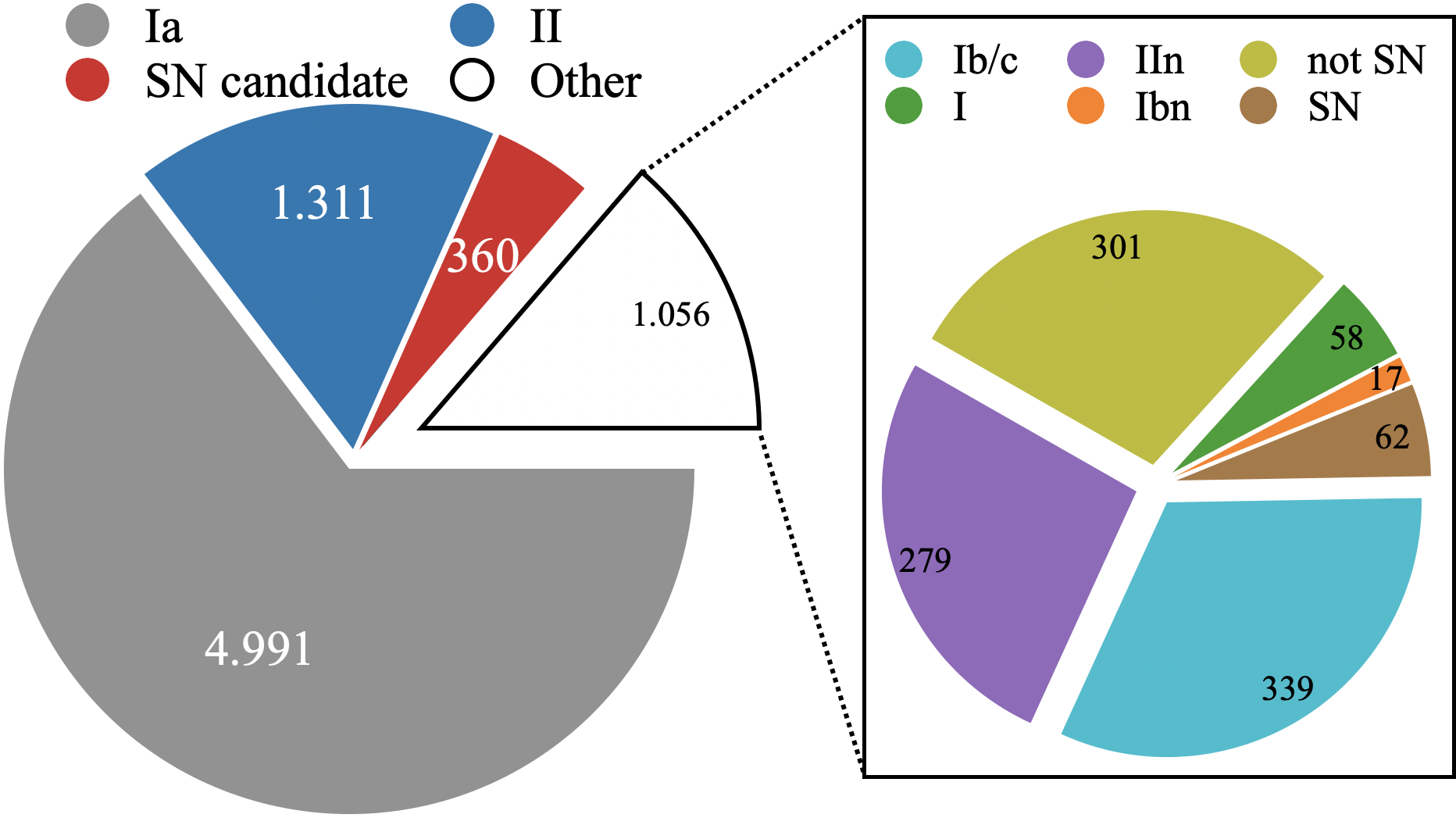}
    \caption{The final sample split into nine classes. As there is a big difference in class sizes, the smaller classes have been put together in the left chart as `other' and are split up in the right chart.}
    \label{class_breakdown}
\end{figure}

We group the transients in our sample into one of nine classes based on the most precise common denominator of the reported classifications. For example, an object that was reported as both a SN Ia and SN Ib is grouped as a `SN I', and an object that was reported as some type of SN I and SN II is grouped as a `SN'. SLSNe are handled the same and grouped with normal SNe into SN I or SN II, though we can recover their SLSN classification if the object proves interesting later. Objects whose classification included the interacting classes SNe Ibn and SNe IIn are kept separately as we expect these may have a higher chance of having a late-time signature that can be picked up by ZTF. Figure~\ref{class_breakdown} shows the nine classes in our sample (`Ia', `II', `Ib/c', `IIn', `Ibn', `I', `SN', `SN candidate', and `not SN') and their relative sizes after a cut on the number of points used in the baseline correction (see Section~\ref{sec:Pipeline_modifications}). 

\begin{figure*}
    \centering
    \includegraphics[width=\textwidth]{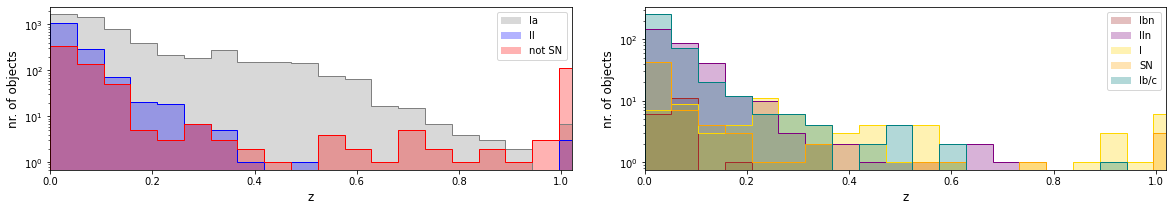}
    \caption{Sample size as a function of redshift for each class of objects. All objects with $z>1$ are put together in the bin starting at $z=1$. The histograms are split into two plots for better readability. In all classes of objects our sample is biased towards lower $z$.}
    \label{class_hists}    
\end{figure*}

For completeness, we include a `not SN' class that consists of other types of transients. This class consists of 60 variable stars which include cataclysmic variables (CVs), luminous blue variables (LBVs), and novae, 36 nuclear transients (including TDEs and AGN), 73 other transients (including gap transients, impostor-SNe, kilonovae), and 140 Long Gamma-Ray Bursts (LGRBs). Figure \ref{class_hists} shows the amount of objects in each class as a function of redshift. Our sample is biased towards lower redshifts as it is magnitude-limited. We also have more objects that have been discovered in years where large surveys such as PTF, iPTF and ATLAS were active.

\section{Analysis}
\label{analysis}
We use an adapted version of the pipeline introduced in \citet{Terwel_2024_paper1} to test for late-time flux excesses in our sample of pre-ZTF transients. In brief, the pipeline first applies a baseline correction to ensure that the light curve has zero flux when no signal is expected \citep[e.g.][]{Yao_baseline_corr, Miller_baseline_corr}. It then bins the post-SN observations together in bins of 25, 50, 75, or 100 days to recover signals that are below the noise level of individual observations. To test if bin placement has an effect on the result, the binning is performed multiple times with shifted bin phase locations. Only signals that are sufficiently insensitive to bin placement are considered real. Binning observations increases the depth at which signals can be recovered at the cost of time sensitivity. As our smallest bins are 25 days in the observer frame, this means that we cannot detect details at similar or smaller time scales, such as rise times or short-term variations. In Section \ref{sec:Pipeline_modifications}, we describe the modifications of our detection pipeline from \cite{Terwel_2024_paper1} and in Section \ref{sec:false_positives}, we discuss the identification of false positives.

\subsection{Detection pipeline modifications}
\label{sec:Pipeline_modifications}
Our main modifications to the pipeline presented in  \citet{Terwel_2024_paper1} are in the method of the baseline correction and the removal of the tail fitting procedure that was put in place to remove false positives from detecting the end of a normal SN Ia tail decline. In \citet{Terwel_2024_paper1} the baseline correction was done by using the pre-SN observations, but this cannot be done in our current sample since the date of explosion is not contained within the ZTF data time frame. Instead, we have to consider that a period with flux excess could occur at any time during the ZTF observations. 

Therefore, we run the pipeline three times, using different time frames each time for the baseline corrections. These time frames are chosen at the start, middle, and end of the ZTF survey span considered (see Table~\ref{baseline_regions}). We choose the baseline time frames to each be a year long to ensure that the targets are likely to have been observed, even if there are gaps in the observations due to the target location not being observable the entire year. When putting the baseline region at the end of the data we make it slightly longer to account for objects having different amounts of data due to their light curves being generated on different days. We also choose not to include the first few months of ZTF data in the first baseline region as the final calibrations were not completed until July 2018 \citep{ZTF_overview_and_1st_results}. By using different baseline regions, we ensure that we bin the entire light curve multiple times (since the region used for the baseline corrections cannot be included in the light-curve binning). If one of the baseline regions overlaps with the late-time signal, the result can be a false baseline correction. By using multiple baseline regions, these cases can more easily be identified. 

\begin{table}[]
    \centering
    \caption{Details of the three baseline regions used in our pipeline.}
    \begin{tabular}{ccccc}
        \hline
        \hline
        Baseline region & Length & \multicolumn{3}{c}{No. passing cuts} \\
        (MJD)& (d) & \ztfg & \ztfr & \ztfi \\
        \hline
        58300 -- 58664 & 365 & 6817 & 7198 & 744\\
        59031 -- 59395 & 365 & 7251 & 7258 & 3447\\
        59915 -- 60335 & 422$^*$ & 4792 & 5350 & 2457\\
        \hline
    \end{tabular}
    \label{baseline_regions}
    \tablefoot{
    The first column gives the start and end MJD of each baseline region, and the second column gives its length in days. The last three columns give the number of transients that had at least 30 points in the baseline to provide a robust estimate of the baseline, in the \textit{g}, \textit{r}, and \textit{i}-band, respectively.\\
    \tablefoottext{*}{The light curves were generated between 9 December 2023 and 24 January 2024, meaning that the final baseline for each object is between 374 and 420 days long. The final two days are a buffer to ensure all data is used.}
    }
\end{table}

To ensure that there were enough points in each baseline region for a robust correction, we ignore all detections where the band in which the detection was found had $< 30$ points in the baseline time frame of approximately one year. With the best ZTF cadence of one data point every two days, the 30 detections cutoff translates to requiring two months of good observations. On the other hand, if the 30 required observations are spread over the entire baseline time frame of 365 -- 422 days (see Table ~\ref{baseline_regions}), the requirement translates into averaging one observation every 12--14 days. Both give a good estimate of the baseline, either by monitoring it closely over a shorter duration or more globally over a longer time period. Table \ref{baseline_regions} shows the number of objects that meet this condition within each baseline region in each of the observational bands. In total, 989 objects never meet this condition in any band and are effectively removed from our sample, reducing the sample to 7\,718 objects.

\subsection{Removing false positives}
\label{sec:false_positives}
The pipeline outputs a list of 360 objects that have $5\sigma$ or greater binned detections in a band in at least four out of 16 attempts (four bin sizes, each shifted four times to avoid spurious detections caused by specific bin placement, see \citealt{Terwel_2024_paper1} for more details). These are inspected visually to determine if the detections are due to observational issues, software issues, or if it is likely astrophysical in nature. A large fraction of the flagged light curves were deemed false positives after visual inspection. The main causes of the binning program wrongly finding bins with detections are due to issues in the difference imaging processing or the baseline correction.

To investigate these cases, we use the SuperNova Animation Program (\textsc{snap})\,\footnote{\url{https://github.com/JTerwel/SuperNova_Animation_Program}} to inspect the difference images directly. Details on \textsc{snap} are presented in \citet{Terwel_2024_paper1}. Many false positives are identified as due to being close to another source by inspection of the difference images. Usually this other source is the host galaxy (nucleus) but in some cases it can also be a foreground star. This can lead to an issue with an improper subtraction of the bright source, which results in a residual, generally a dipole, at the position that is picked up by our pipeline. To identify these cases, we use \textsc{snap} to visually inspect the images and remove spurious detections. This removes 155  events from our sample. 

The other main group of false positives (88 cases) are flagged due to issues in the baseline determination. These can present as extremely large corrections, in some cases several orders of magnitude larger than the signals we expect to find. A baseline correction of $\mathcal{O}(10^5)$ can make a 17.5 mag signal appear or disappear. While in some cases big corrections are to recover fading transients that were present in the reference image (see Section~\ref{tails}), in most cases such a correction led to noisy light curves, preventing us from probing beyond the individual ZTF image mag limit. A failure during image processing or an incorrectly estimated baseline correction can also result in large corrections. In other cases the baseline is not constant but seems to vary over time or suddenly jump, making it very difficult or impossible to apply a proper baseline correction.

\begin{figure*}
    \centering
    \includegraphics[width=\textwidth]{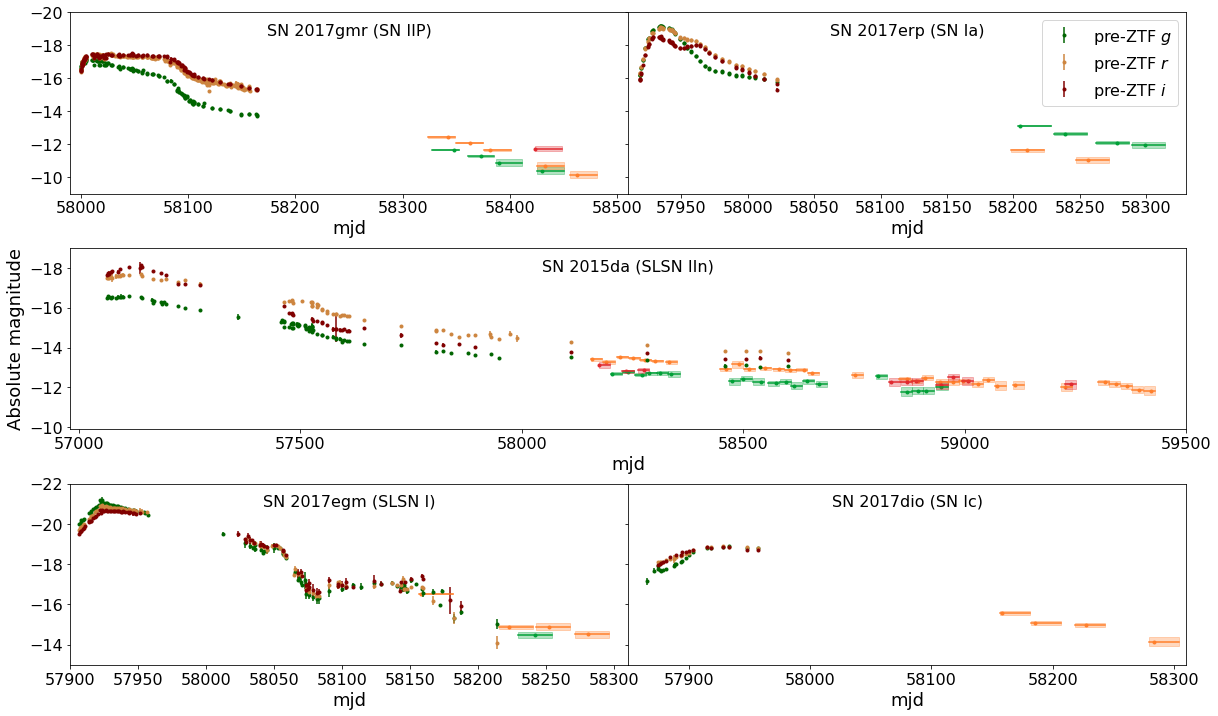}
    \caption{Examples of pre-ZTF SNe whose light curves have been recovered in the binned ZTF light curves. For each object we show the pre-ZTF \ztfg~(dark green), \ztfr~(dark yellow), and \ztfi~(dark red) data points, as well as the binned ZTF detections (green, orange, red, respectively). The binned ZTF observations are in 25-day bins, with the width showing the beginning and end of each bin, and the shaded region shows its $1\sigma$ magnitude uncertainty. The pre-ZTF data was taken from \citet{2017gmr} (SN 2017gmr), \citet{2017erp} (SN 2017erp), \citet{2015da_2020} (SN 2015da), \citet{2017egm} (SN 2017egm), and \citet{2017dio} (SN 2017dio). None of the light curves are corrected for host extinction.}
    \label{tail-examples}
\end{figure*}

One potential cause of these issues could be due to ZTF continuing to update the reference images by rebuilding them and stacking more observations, including observations from during the survey \citep{ZTF_Instrumentation}. To be able to compare observations from before and after this is done, one should remake all the difference images using the updated references. However, this is not feasible in a survey as large as ZTF, and since the offsets are usually below the noise threshold of the individual images the issue is of little importance to most of the survey science outputs. Only when attempting to go beyond the single image noise limit using, for instance, our binning method, this issue becomes noticeable enough and can lead to baseline offsets and varying or jumping baselines.

In 19 cases there were enough points in the baseline for the object to stay in our sample but the light curve was sampled sparsely with large gaps without observations surrounding sparse detections, making it impossible to determine the validity of these detections as there are no proper non-detections close in time to compare against. These cases were removed along with the other false positives.

\section{Results}
\label{results}
We have identified 98 transients with potential late-time excesses in their light curves that require further investigation. In Section~\ref{tails}, we describe the 63 objects whose detections in ZTF can directly be linked to the pre-ZTF transient. These transients were still bright enough at the start of the ZTF survey to be detectable. In Section~\ref{siblings}, we describe 12 objects whose found ZTF signal is due to a sibling transient occurring at nearly the exact same sky position. A small fraction (14 objects) of our sample consists of non-SN transients and are discussed in Section~\ref{non-sn}. Finally, in Section~\ref{weirdo_section}, we describe the nine objects that required an individual, deeper investigation of the signal found in ZTF.

\subsection{Pre-ZTF transients still active in ZTF}
\label{tails}
We chose to limit our sample to those objects that were first detected before 2018 to reduce the number of transients still visible at the start of ZTF. While this three-month gap is enough for most transients to fade away, some super-luminous SNe, Type II-Plateau SNe, interacting classes like SNe Ibn and IIn, and even very nearby SNe Ia that exploded before ZTF started, may still be active at the start of ZTF.

As these objects were active while the initial set of ZTF reference images were being produced, none of these objects have been found by ZTF even when the SN was still bright enough to be detected by ZTF once observing began. As the transient is in the reference images, it will cause an over-subtraction and leave an imprint, or ghost, at its location in the difference images. These are easily recognisable through visual inspection of the difference images using \textsc{snap}. Another clear sign is a significant baseline correction that is consistent between the different baseline regions after the transient has faded away. The baseline correction corrects for the flux offset created by the ghost, revealing the tail as observed by ZTF in the light curve.

\begin{figure*}
    \centering
    \includegraphics[width=\textwidth]{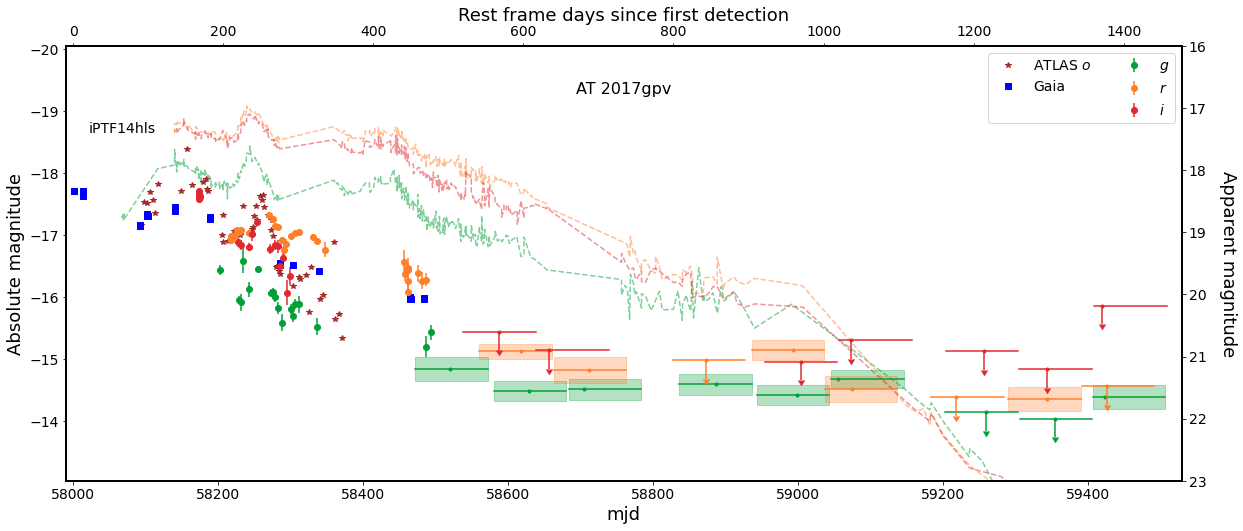}
    \caption{Light curve of AT 2017gpv with the axes representing absolute magnitude (left), apparent magnitude (right), the rest frame days since first detection (top) and mjd (bottom). Single epoch detections by ZTF, Gaia, and ATLAS are shown, as well as the binned ZTF observations after the transient faded below the single epoch noise limit. The dashed lines are the \textit{gri}-band light curves of iPTF2014hls, which have been corrected for time dilation but not extinction (data taken from \citealt{iPTF14hls_Iair, Sollerman_2019_iptf14hls}).}
    \label{17gpv_plot}
\end{figure*}

We find 63 transients whose ZTF detections are consistent with ongoing transient flux. These are listed in Table~\ref{tail_objects} and some example light curves are shown in Fig.~\ref{tail-examples}. The light curves show pre-ZTF data taken from the literature for  SN 2017gmr \cite[][]{2017gmr},  SN 2017erp \cite[][]{2017erp}, SN 2015da \cite[][]{2015da_2020}, SN 2017egm \cite[][]{2017egm}, and SN 2017dio \cite[][]{2017dio}. SN 2015da is an extremely slowly declining SLSN IIn event \citep{2015da_2020, 2015da_2024} and its light curve extends in the binned ZTF data to approximately eight years after discovery. In SN 2015da, the pre-ZTF and binned ZTF light curves do not overlap exactly in magnitude at epochs when data from the different surveys are available (MJD 58300 -- 58600), showing that our pipeline underestimates the brightness of the transient. This is because some SN flux is still present even in our latest baseline region during mainly 2023, as can be seen in the observations presented in \citet{2015da_2024}. Therefore, the baseline correction is too small causing the binned detections to be slightly too low.

\subsubsection*{AT 2017gpv - a 14hls-like event}
AT 2017gpv, shown in Fig.~\ref{17gpv_plot}, is an unusual event similar in nature to iPTF14hls \citep{iPTF14hls_Iair, Sollerman_2019_iptf14hls} and SN 2020faa \citep{Yang_2021_20faa, 2020faa_hidden_shocks} that was detected by our pipeline in the ZTF data. It was originally identified by Gaia, which detected it repeatedly during the first 500 days after its discovery. It has also been detected by ATLAS, which has a rich $o$-band light curve between 100 and 300 days after discovery that shows a plateau followed by a shallow decline that is interrupted by a rebrightening event. After the baseline correction, the first detections in ZTF are visible even in the non-binned data, beginning around 200 days after first detection and lasting for around 400 days. With the binned observations, the object can be recovered for another 1\,000 days before it fades below the noise limit.

Due to the long duration of detections of AT 2017gpv in ZTF, only the last baseline region can be trusted as it has the least contribution from the late-time signal, though given the slow decline of the transient it is likely that there could still be some excess present at the time of the last baseline region. As expected, the baseline corrections are significant, and \textsc{snap} clearly shows the transient in the reference images, as well as a significant ghost in all difference images. While it was picked up by the ZTF alert system and given an internal name (ZTF18acueiall), it was not recognized as a real transient.

Unfortunately, AT 2017gpv was never spectroscopically classified. The transient is at a distance of 6.88\arcsec\ from the host nucleus, removing host variability as a possible explanation, as well as limiting the amount of host extinction expected. The long time it was detectable and the bumpy nature of its light curves look similar to iPTF14hls \citep{iPTF14hls_Iair, Sollerman_2019_iptf14hls}, a very peculiar SN II that is also shown in Fig.~\ref{17gpv_plot}. Some similar events to iPTF14hls have been identified, but they are rare \citep{Yang_2021_20faa, Soraisam_2022}. The light curve of iPTF14hls spans over 600 days and is very bumpy. In both cases the explosion epoch is badly constrained, but the peak found in iPTF14hls matches up quite well with the bump around MJD 58250 in AT 2017gpv. Overall, AT 2017gpv looks like a fainter and somewhat faster decaying version of iPTF14hls. The late-time data of iPTF14hls \citep{Sollerman_2019_iptf14hls} extends to $\sim1\,200$ d after discovery and shows a sharp decline after $\sim1\,000$ d, which is not seen in AT 2017gpv.

iPTF14hls is in our initial sample but was not detected in ZTF because it is $\sim1\,000$ days older than AT 2017gpv. Assuming that these two events evolved similarly, iPTF14hls would have been close to the detection limit at the start of ZTF and faded below that before it could be picked up by our pipeline.

\begin{table*}
    \centering
    \caption{Pre-ZTF transients with a sibling transient detected in ZTF in single exposures.}
    \resizebox{\textwidth}{!}{
    \begin{tabular}{cccc|cccccc}
        \hline
        \hline
        \multicolumn{4}{c|}{Pre-ZTF transient} & \multicolumn{6}{c}{ZTF transient}\\
        Name & Type & MJD & Redshift & ZTF name & IAU name & Type & MJD & Peak mag & Sep (\arcsec)\\
        \hline
        AT 2017gcd & ?$^*$ & 57968 &0.028593 & ZTF19acdgwhq & SN 2019aavr & Ia-norm & 58763 & $-$18.8 & 1.02\\
        AT 2017keg & ?$^*$ & 58068 & 0.05 & ZTF19acihgng & SN 2019tka & Ia-norm & 58781 & $-$19.1 $^g$ & 1.69 \\
        SN 2013ld & ?$^*$ & 56522 & 0.02741 & ZTF18abavruc & SN 2021rgw & Ia & 59393 & $-$18.9 $^g$ & 0.57 \\
        ASASSN-14ba & Ia pec/91T & 56796 & 0.032668 & ZTF22aaawghw & AT 2022csd & II & 59629 & $-$16.8 & 1.29 \\
        SN 2017acp & II & 57785 & 0.0215 & ZTF19aavqics & SN 2019gxo & II & 58633 & $-$18.1 & 3.15 \\
         SN 2014gz & II & 56678 & 0.02558 & ZTF21abcpbqd & SN 2021nof & II & 59362 & $-$17.2 $^i$ & 0.30 \\
        SN 2009hz & II & 55046 & 0.0253 & ZTF18acotwcs & AT 2018iml & CV?$^{**}$ & 58439 & -- & 0.33\\
        iPTF15wk & Ia & 57097 & 0.23 & ZTF21aaeebxm & - & ? & 59226 & $-$20.0 $^g$ & 0.31 \\
        SN 2016bsc & Ia & 57500 & 0.05 & ZTF23aaeljse & (SN 2016bsc) & ? & 60043 & $-$17.9 & 1.61 \\
        SNF20080522-001 & Ia & 54608 & 0.04872 & ZTF22aalbuig & AT 2022kuh & ? & 59724 & $-$17.3 & 1.64 \\
        PS15ctg & Ia & 57328 & 0.078 & ZTF22abamjrf & AT 2022rol & ? & 59806 & $-$18.7 & 1.03 \\
        iPTF15eot & II & 57357 & 0.039 & ZTF19aakvysq & AT 2019bll & ? & 58541 & $-$17.5 & 2.19 \\
        \hline
    \end{tabular}
    }
    \tablefoot{
        The first four columns give the name, type, discovery mjd, and redshift of the pre-ZTF transient, and the next five columns give the same information about its ZTF sibling when available. The second to last column gives an estimate of the peak \ztfr-band (unless specified otherwise) absolute magnitude of the ZTF transient assuming the pre-ZTF transient redshift, except for AT 2018iml as this is likely a foreground CV. The last column gives the separation between the siblings. A ? type means the transient was never spectroscopically classified. IAU names in parentheses are ZTF transients wrongly associated with a pre-ZTF transient.\\
        \tablefoottext{*}{Classified ZTF SNe that were wrongfully associated with the pre-ZTF name resulting in a misclassification. The pre-ZTF transient itself was never actually classified in the cases of AT 2017keg and SN 2013ld. The case of AT 2017gcd / SN 2019aavr has been fixed on TNS.}\\
        \tablefoottext{**}{Not officially classified but based on evidence gathered from the ZTF forced photometry light curve.}\\
        \tablefoottext{g}{Peak absolute magnitude in the \ztfg-band.}\\
        \tablefoottext{i}{Peak absolute magnitude in the \ztfi-band.}
        }
    \label{sibling_table}
\end{table*}

\subsection{Siblings}
\label{siblings}
Siblings are two (or more) transients that occur in the same host galaxy. While siblings can occur at any location in a galaxy \citep[see e.g.][for a sample of ZTF-detected siblings]{BTS_siblings, DR2_siblings}, a subset of these occur with a small enough sky separation that part of the light of one sibling can be detected when performing forced photometry at the sky position of the other sibling. In \citet{Terwel_2024_paper1}, five such sibling pairs were found, with both transients detected within ZTF. With our current sample being larger and spanning a bigger time range over which the second transient can be observed, it is reasonable to expect a larger number of same-location sibling transients. We arbitrarily define siblings as transients detected in ZTF without the use of additional binning to push the detection limit, that are distinctively separate in time from their pre-ZTF transient sibling counterpart and at a separation small enough to be picked up in forced photometry centred at the location of the pre-ZTF transients. The ZTF transients do not have to be classified. We find 12 pairs of transients that satisfy these conditions, which we verified through the Fritz broker \citep{skyportal2019, Skyportal}. These are shown in Table~\ref{sibling_table}. 

In three cases, the ZTF-detected transient was classified but mistakenly associated with a pre-ZTF transient in WISeREP or OSC that subsequently got the same classification. In the case of AT 2017gcd, a decaying transient was observed in four epochs of unforced Pan-STARRS photometry spread out over 125 days. This is enough to conclude that the 2017 transient was real and likely some kind of SN.

In the two other cases (AT 2017keg/SN 2019tka and SN 2013ld/SN 2021rgw), the pre-ZTF detections are spurious and are unlikely to be true sibling pairs. AT 2017keg was reported by ATLAS in 2019 \citep{2017keg_disc} when SN 2019tka was found. Due to SN 2019tka being close to the host nucleus, likely spurious detections from two years before were present in the ATLAS light curve, resulting in the automated discovery report stating the wrong discovery date. SN 2019tka was reported by ZTF \citep{2019tka_disc}, which did not have such earlier detections as it only started operating in 2018. SN 2013ld was reported in 2021 by Pan-STARRS1. Again, its sibling SN 2021rgw was on top of the host nucleus, which has had several epochs of minor variability. Even the internal ZTF name is from 2018, showing that ZTF also detected minor changes at the host nucleus location. When SN 2021rgw was found, ZTF issued an alert with the discovery date in 2021, resulting in SN 2021rgw \citep{2021rgw_disc} while Pan-STARRS1 used the first unforced detection epoch in 2013 as the discovery date, resulting in SN 2013ld \citep{2013ld_disc}.

For the other nine sibling pairs listed in Table \ref{sibling_table}, the pre-ZTF sibling has a classification (five SNe Ia and four Type II SNe) obtained at the time of discovery. Three of their paired ZTF siblings were spectroscopically classified as Type II SNe at the time of discovery in ZTF. For AT 2018iml, the sibling of Type II SN 2009hz, the ZTF detections were obtained in two periods (November 2018 and March 2024) and they are best matched to a CV. The remaining five had no ZTF-era classification. In some cases the ZTF sibling is close enough to get wrongly associated with the pre-ZTF sibling and obtain the same IAU name.

Assuming the five unclassified ZTF transients occurred in the same host galaxy as their classified sibling, we can use the known redshifts to estimate the absolute brightness of the unclassified transients and show that they are in the range of SNe and unlikely to be caused by late-time CSM interaction.
ZTF21aaeebxm is found around 1\,800 days after its sibling, iPTF15wk, was first detected, and has an inferred absolute \textit{g}-band magnitude of $-20.0$ at its peak at just 0.31\arcsec\ from iPTF15wk. To get such a strong signal this long after the explosion would require an unreasonably large CSM and thin shell, making CSM interaction an unlikely explanation. Its location is consistent with its host nucleus, making a nuclear transient origin likely.

For ZTF23aaeljse, ZTF22aalbuig, ZTF22abamjrf, and ZTF19aakvysq, their absolute peak magnitudes are in the range typical of SNe ($-$17.3 to $-$18.7 mag), and \textsc{snap} shows that there is a noticeable separation between the pre-ZTF and ZTF-detected signals (1.03 to 2.19\arcsec), disfavouring any explanation that would require the two events to be at the same spatial position, such as CSM interaction. The inferred CSM masses to explain these events in this way are also unrealistically high. Their locations are inconsistent with their host nuclei. Therefore, we conclude that a sibling transient, likely an unclassified SN, is the most plausible explanation for these events.

\subsection{Non-SN sources of flux excesses in ZTF}
\label{non-sn}
\begin{table}
    \centering
    \caption{Pre-ZTF transients detected in the binned ZTF light curves due to AGN or variable star activity.}
    \resizebox{\columnwidth}{!}{
    \begin{tabular}{cccc}
        \hline
        \hline
        Name & Pre-ZTF type & $z$ & Late-time type\\
        \hline
        Gaia14adg & II & 0.154 & AGN\\
        SN 2016fiz & II & 0.05 & AGN\\
        SN 2017avb & II & 0.096 & AGN\\
        LSQ12biu & IIn & 0.136 & AGN\\
        SN 2017bcc & SLSN-II & 0.133 & AGN\\
        PSN J0151 $^1$ & SLSN-II?/AGN & 0.26 & AGN\\
        ATLAS17khl & AGN & 0.06 & AGN\\
        PS17bgm & AGN & 0.358 & AGN\\
        LSQ12ehj & AGN & 0.12 & AGN\\
        AT 2017kas & SN candidate & 0.031328 & AGN\\
        LSQ12fgx & Variable star & 0 & Variable star\\
        SNhunt44 & LRV? $^2$ & 0.0005864 & Variable star\\
        AT 2016ijb & SN candidate & $-$0.000781 & Variable star\\
        PTF10qpf & Variable star & 0 & Variable star\\
        \hline
    \end{tabular}
    }
    \label{non-transient_table}
    \tablefoot{
    The original type and $z$ values are as they were recorded on the OSC or WISeREP. The late-time type gives the reason for the ZTF detections.\\
    \tablefoottext{1}{PSN J0151 = PSN J01510869+3155215\\}
    \tablefoottext{2}{Potential luminous red variable}
    }
\end{table}

\begin{table*}
    \centering
    \caption{Objects with detections whose origin was not immediately clear.}
    \resizebox{\textwidth}{!}{
    \begin{tabular}{ccccc|cccccccccc}
        \hline
        \hline
        \multicolumn{5}{c}{Pre-ZTF} & \multicolumn{6}{|c}{ZTF excess}\\
        \hline
        Name & Type & Discovery & $z$ & Host & Start & Duration & \ztfg\ band & \ztfr\ band & \ztfi\ band & Inferred & \multicolumn{2}{c}{Excess consistent with}\\
        && date && sep (\arcsec) & (d) & (d) & abs. mag & abs. mag & abs. mag & cause of excess & SN & host nucleus \\
        \hline
        SN 2017ige $^1$ &Ia & 17-11-17 & 0.02431 & 0.99 & 1130 $^2$ & 90 & $-$13.5 -- $-$13.3 & $-$14.3 -- $-$13.6 & -- & Sibling& yes & no \\
        SN 2016cob & Ia-91T & 26-05-16 & 0.02961 & 0.7 & 2080 & 480 $^3$ & $-$15.3 -- $-$14.7 & $-$16.5 -- $-$16.1 & $-$17.2 -- $-$16.4 & CSM/Nuclear & yes &yes\\
        SN 2017frh & Ia & 17-07-17 & 0.032188 & 0.0 & 230 $^2$ & 180 $^3$ & $-$15.6 -- $-$14.9 & $-$16.0 -- $-$15.6 & -- & CSM/Nuclear & yes & yes \\
        SN 2017fby & Ia & 01-07-17 & 0.043513 & 0.73 & 1780 & 190 $^3$ & -- & $-$16.5 -- $-$16.1 & -- & CSM/Nuclear & yes & yes \\
        PTF10jtp & Ia & 04-06-10 & 0.067 & 1.11 & 2670 & 200 $^3$ & -- & $-$16.5 -- $-$16.3 & -- & Nuclear& yes & yes \\
        AT 2017fwf & Cand. & 01-08-17 & 0.033707 & 1.67 & 1690 & 410 & -- & $-$14.7 -- $-$14.0 & -- & Nuclear& no & yes \\
        PSc130283 & Ia & 29-01-11 & 0.07622 & 0.44 & 3750 & 480 $^3$ & -- & $-$17.8 -- $-$16.9 & -- & Nuclear& yes & yes \\
        PTF13cow & Ia & 07-08-13 & 0.086 & 0.56 & 2980 & 560 $^3$ & $-$16.3 -- $-$16.1 & $-$17.4 -- $-$17.1 & -- & Nuclear& no & yes \\
        iPTF15aow & Ia & 06-05-15 & 0.07597 & 0.92 & 1970 & 1050 $^3$ & $-$16.0 -- $-$15.9 & $-$17.2 -- $-$16.4 & $-$18.0 -- $-$17.1 & Nuclear& yes & yes \\
        \hline
    \end{tabular}
    }
    \label{weirdos}
    \tablefoot{
    The first five columns give information on the original transient, and the last six columns give an overview of the binned ZTF detections. Start gives the time after discovery of the first detections in ZTF, and duration gives the length of these detections. Both are in rest frame days and rounded to 10 days. The \ztfg, \ztfr, and \ztfi~columns give the range of detections in absolute magnitude if detected. Inferred cause of excess gives the type of object that best explains the detections. The last two columns state whether the excess is consistent with the SN and host nucleus location, respectively.\\
    \tablefoottext{1}{SN 2017ige has two separate periods with detections. The second is shown in the table while the first is consistent with the radioactive tail phase of the SN Ia. It starts at 80 d after discovery (the start of the ZTF survey) and lasts for 150 d declining from $-$14.6 to $-$13.8 mag over this time in the \textit{g} band.\\}
    \tablefoottext{2}{Detected from the start of ZTF.\\}
    \tablefoottext{3}{Start or end with a gap in the binned observations or at the edge of available data.}
    }
\end{table*}

\begin{figure*}
    \centering
    \includegraphics[width=0.95\textwidth]{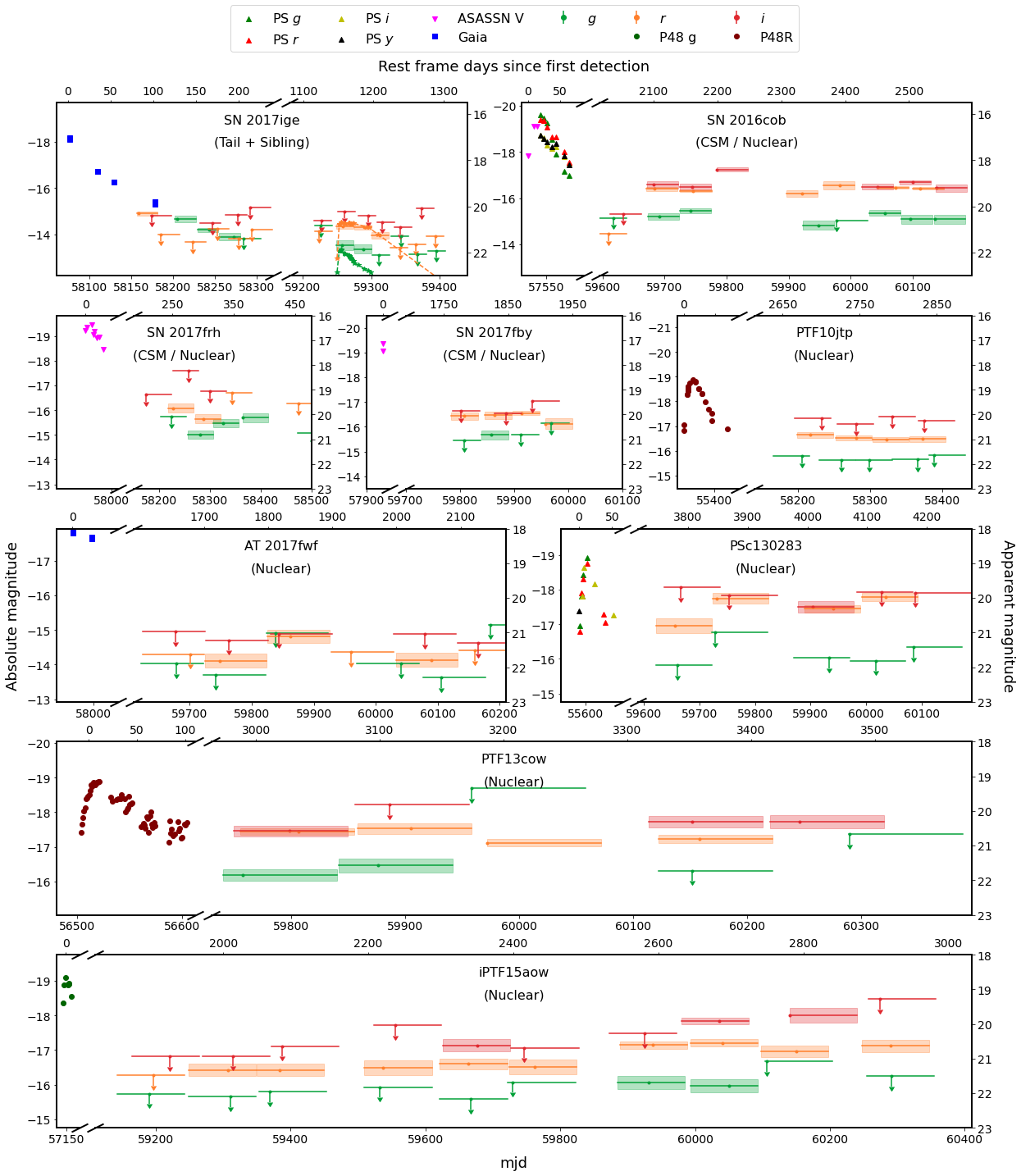}
    \caption{The nine objects whose late-time detections required a more rigorous investigation. Each object is shown in absolute and apparent magnitude (left and right axes respectively) and rest frame days since first detection and mjd (top and bottom axes respectively). We show the binned ZTF data with the late-time detections as well as pre-ZTF detections from other surveys (Pan-STARRS, Gaia, ASASSN, PTF) where available. For SN 2017ige a heavily extinct (E(B-V)$_\text{host}=1.2$ mag) light curve of SN 2020jfo is shown with dashed lines matching the late-time ZTF $r$-band detections. The objects are marked with the explanation for their late-time detections.}
    \label{weirdo_plots}
\end{figure*}

We identify some late-time excesses at the positions of our sample that are astrophysical but are due to known AGN activity or stellar variability. Ten transients (listed in Table~\ref{non-transient_table}) in our sample have been found to have genuine long-term variability due to an AGN that lasts for the whole time period of ZTF and is picked up by our pipeline. In each case, the source is a known AGN (verified through cross-referencing with SIMBAD, ZTF, and checking the AGN criterion from \citealt{WISE_crit} using the Wide-field Infrared Survey Explorer WISE \citealt{WISE}). Their light curves are shown in Fig.~\ref{non-transients_AGN}. While several of the pre-ZTF transients in this category have been classified as SNe II, some have high redshift, which suggests that these are AGN misclassified as SNe II.

We also recovered four long-term variable stars that are shown in Fig.~\ref{non-transients_varstar} and listed in Table~\ref{non-transient_table}. The nature of these sources makes it impossible to create a template that always subtracts them completely without leaving residual flux or a ghost as there may be no region without variability. However, if the variations in the source's magnitude are large enough over a long period of time, it will easily be picked up by the binning algorithm.

\subsection{Final shortlist of late-time interaction}
\label{weirdo_section}
The final nine objects in our sample cannot be put in any of the groups (ongoing SN flux, sibling transient, nuclear activity, variable star) above. Each of these events is discussed individually below. Table \ref{weirdos} shows the general information of these events, and their light curves are shown in Fig.~\ref{weirdo_plots}. The bin sizes and placement chosen for the plots are those that result in the clearest and cleanest signals. The baseline regions used are those that have the least amount of transient flux in them, which means they are the furthest away in time from the excess. As was done in \citet{Terwel_2024_paper1}, we compare the ZTF detections to several classes of transients in an attempt to explain them as a previously unidentified sibling transient. We also estimate the potential CSM mass required to explain the late-time signature using the analysis method described below for objects where a CSM interpretation cannot be ruled out.

\subsubsection*{Estimating CSM masses}
\label{CSM_calc}
To determine if the objects in our final shortlist are potentially due to CSM interaction signatures, we can make a rough estimate of the CSM mass required to generate such a signal. In this section, we describe the method used by \citet{2015cp} to put constraints on the CSM mass of their targeted SNe at similar epochs to ours, based on the observed near-ultra-violet (NUV) (non-)detections using the models of \citet{CSM_models_Harris}. While CSM is line-dominated, especially by \Halpha, it is much more difficult to estimate a CSM mass without making many assumptions about the state of the CSM. Even though the statistical errors can be large, the resulting CSM masses for some of the objects are still unreasonable.

Assuming that the time of the late-time detections corresponds to the peak of the CSM interaction light curve, \citet{2015cp} showed that the Bremsstrahlung spectral luminosity $L_\nu$ at frequency $\nu$ at the moment the interaction shock reaches the outer edge of the shell is
\begin{equation}
    \label{15cp_L_nu}
    L_\nu \approx 1.63 \times 10^{-31} T^{-1/2} t_r^{-3} M_\text{CSM}^{17/7} \text{e}^{-\frac{h\nu}{kT}} y(F_R) \text{ erg s}^{-1}\text{Hz}^{-1},
\end{equation}
where $T$ is the temperature of the shocked material in Kelvin, $t_r = t/(1+z)$ is the time after explosion in seconds in the rest frame of the SN at which the ejecta reach the outer edge of the CSM shell and the interaction is assumed to be at its strongest, $M_\text{CSM}$ is the CSM mass in grams, and $y(F_R) = F_R^{-3/7}(1-F_R^{-3})^{-10/7}$ is the dependence on the fractional radius of the shell $F_R \equiv R_\text{out} / R_\text{in}$. Here $R_\text{out}$ and $R_\text{in}$ are the outer and inner radius of the CSM shell, respectively. $F_R = 1.1$ represents a thin nova-like shell, with higher values representing thicker shells. This equation assumes a low-density, fully ionized H-dominated CSM to be the emission source after the interaction. 

\citet{2015cp} derive Eq.~\ref{15cp_L_nu} to use in the NUV and assume temperatures around $10^8$~K. In these conditions the exponential term is near unity and can be ignored. Since we are targeting optical wavelengths, if the mechanism described here is the source, lower temperatures are needed where the exponential term cannot be ignored. The other assumptions made in \citet{2015cp} still hold for optical wavelengths, so we can use Eq.~\ref{15cp_L_nu} to estimate the $M_{CSM}$ required to explain any detected late-time signals using
\begin{equation}
    \label{mag_lum_rel}
    10^{-0.4M} = \frac{L}{L_0} = \frac{L_\nu c}{L_0 \lambda},
\end{equation}
with $M$ the absolute magnitude of the detected signal, $c$ the speed of light, $L_0 = 3.0128 \times 10^{35}$ erg s$^{-1}$ the zero point luminosity, and $\lambda$ the effective wavelength of the band we are considering ($4\,746.48$ \AA, $6\,366.38$ \AA, and $7\,829.03$ \AA~for \textit{g}, \textit{r}, and \textit{i}, respectively). By combining Eq.~\ref{15cp_L_nu} and Eq.~\ref{mag_lum_rel} we get
\begin{equation}
    \frac{M_\text{CSM}}{M_\odot} \approx 3 \times 10^{-8-\frac{14}{85}M} \left(\frac{T}{\text{K}}\right)^{\frac{7}{34}} \left(\frac{t/(1+z)}{\text{days}}\right)^{\frac{21}{7}} \left(\frac{\lambda}{\text{\AA}}\right)^{\frac{7}{17}} \text{e}^{\frac{7hc}{17kT\lambda}} y(F_R)^{\frac{-7}{17}}.
    \label{M_CSM_eq}
\end{equation}
We can obtain $M$, $t$, and $z$ directly from our shortlist of transients with potential CSM interaction. For $t$ the start of the signal should be taken as this assumes the least delay due to a non-negligible light-crossing time across the CSM shell. The remaining parameters are the fractional radius of the shell and the temperature.

$F_R=1.1$ if we assume the CSM shell to be nova-like. Equation~\ref{M_CSM_eq} easily gives very massive CSM shells, especially if there is a significant delay before the onset of the CSM interaction. The required $M_\text{CSM}$ can be somewhat lowered if we assume that the actual thickness $R_\text{in} - R_\text{out}$ of the CSM shell remained constant after its creation, meaning that $F_R$ decreases as the shell travels outwards. Since the SN ejecta move at a constant velocity, the distance of the CSM is proportional to the delay time of the interaction signal. Assuming a shell has $F_R = 1.1$ if the late-time signal occurs 300 days after the explosion, that same shell will have $F_R \approx 1.01$ if the interaction starts 3\,000 days after the explosion. The main reason for allowing $F_R < 1.1$ is to see how thin a shell would need to be to give a reasonable $M_\text{CSM}$. We assume temperatures of 10$^5$ to 10$^7$ K, lower than the value assumed by \cite{2015cp} of 10$^8$ K, but more suitable for explaining optical emission.

Equation~\ref{M_CSM_eq} assumes that the magnitude is corrected for extinction. To get the lowest $M_\text{CSM}$ possible we assume there is only Milky Way extinction in the line of sight of any of our targets. Any extinction in the host galaxy would make the intrinsic colour of these objects bluer allowing for higher temperatures, but also raise the intrinsic absolute magnitude, resulting in a higher overall estimate for $M_\text{CSM}$. The estimates should therefore be seen as lower limits.

\subsubsection*{SN 2017ige}
SN 2017ige was initially discovered and classified as a SN Ia in late 2017, only a few months before the start of ZTF. As it was relatively close by, the radioactive tail phase was identified in the ZTF binned data out to $\sim200$ d, before it faded below the noise limit. This tail matches well with the declining light curve seen in the early ($<100$ d) Gaia data (see Fig.~\ref{weirdo_plots}). \textsc{snap} shows that the SN is in the ZTF reference images, leaving a small ghost at its location in the difference images. 

At 1\,000 days after the SN tail faded below the noise limit, a small excess is detected in both the ZTF \ztfg- and \ztfr-bands that lasts around 90 days. In the difference images a small excess can be seen semi-overlapping the negative imprint left by SN 2017ige during this excess. Such a dipole signal would usually suggest imperfect image subtractions, but in this case it can also be interpreted as a separate transient slightly offset from the location of SN 2017ige being over-subtracted due to the presence of a SN in the reference images.

We use the duration and shape of the identified excess to test if it could be explained as a sibling transient. A light curve similar to that of SN 2020jfo \citep{Sollerman_2020jfo, IIp_ext}, a Type IIP SN with a relatively short plateau, fits the excess quite well in duration and absolute magnitude when a host extinction of $ E(B - V)$ = 1.2 mag is added. Given that the sky position at which the forced photometry was performed is only 0.99\arcsec\ from the host nucleus of an edge-on galaxy, such a high amount of extinction is plausible. The original peak of the light curve of SN 2017ige in 2017 was not caught so a comparison with a similar extinction estimate cannot be made. Although the late-time excess is close to the nucleus, inspection of the images shows that it is offset from the host centre as well as the original SN position. Therefore, we conclude that a Type II SN sibling with a relatively high extinction value is an adequate explanation for this late-time excess.

\begin{figure*}
    \centering
    \includegraphics[width=16cm]{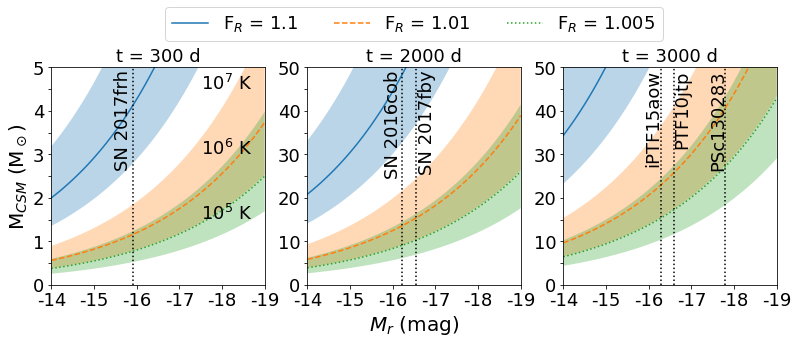}
    \caption{$M_\text{CSM}$ as a function of the \ztfr-band absolute magnitude $M$ for different assumptions of $t$ (panels), $T$ (shaded region), and $F_R$ (colours). The five objects for which we estimate $M_{CSM}$ in Sect.~\ref{weirdo_section} have been marked in the panel that has $t$ the closest to the detected signal. Decreasing shell thickness and temperature to mitigate $M_\text{CSM}$ is effective, though there is a limit to how far this can be done while keeping realistic values for $F_R$.}
    \label{M_CMS_fig}
\end{figure*}

\subsubsection*{SN 2016cob}
SN 2016cob was classified as a SN Ia of sub-class 91T-like at early times. There is light curve data from Pan-STARRS and ASASSN around peak and extending to just past 50 d after the peak. In the binned ZTF data, this event jumps up from non-detections to detections in all three bands around $\sim2\,150$ d after first detection (Fig.~\ref{weirdo_plots}). The \ztfi-band data is included for completeness but does not have a reliable baseline estimate ($>$30 points per region) so the values should be treated with caution. The detections are present in all bands at the same time and extend for at least 480 d at an absolute magnitude of $-$15 to $-$17 mag. No SN-like transient can explain this long-lived flux excess.

Following the calculation in Section \ref{CSM_calc}, if we assume a CSM shell with a fractional radius of $F_R\approx 1.005$ at a temperature of 10$^5$ K, this gives $M_\text{CSM}\approx$ 6 $M_\odot$. In Fig.~\ref{M_CMS_fig} we show the $M_\text{CSM}$ as a function of the \ztfr-band absolute magnitude, $M_r$, for three different times of the onset of the CSM interaction (300, 2\,000, 3\,000 d), three different $F_R$ values (1.1, 1.01, and 1.005) and temperatures of 10$^5$ to 10$^7$ K. A CSM mass of $M_\text{CSM}\approx6M_\odot$ is large, but around the same amount as suggested to explain the interaction signatures seen in SN 2002ic \citep{Hamuy_02ic}. Based on this, we put SN 2016cob forward as a candidate for showing late-time CSM interaction, though we note that it would require a very thin shell ($F_R$ = 1.005).

SN 2016cob is 0.7\arcsec\ from the host nucleus, putting the nucleus on the same or an adjacent pixel in the ZTF observations. This leaves room for a host variability interpretation, though the host has no AGN according to its WISE colours. \textsc{snap} does not show any reduction or subtraction issues that could explain the ZTF detections as a spurious source either.

The absolute magnitude, duration, and nuclear location are consistent with the ambiguous nuclear transient (ANT), ASASSN-20hx/AT 2020ohl \citep{Hinkle_Extreme_nuclear_transients/ANTs}. ANTs are events that cannot be easily classified into AGN activity or TDE \citep{wiseman_ztfants}. AT 2020ohl displayed a plateau or very slow decline in its optical light curves for $>250$ d relative to peak with an approximate plateau magnitude in the \textit{gri}-bands of $-$17.5 mag. In Fig.~\ref{ANT_comp}, we compare the light curves of the shortlisted transients that are consistent with their host nuclei compared to AT 2020ohl. The flux excess at the position of SN 2016cob is shown as blue squares and is slightly fainter than AT 2020ohl. AT 2020ohl had an observed rise-time of 30 days \citep{Hinkle_Extreme_nuclear_transients/ANTs}, slightly larger than our smallest bins. This would explain why no rise is detected, as our time resolution is too poor to detect time variations of this scale. We can only observe a sudden appearance of a very flat light curve. Both late-time CSM interaction and an ANT are adequate explanations for the identified signal, and without additional information we cannot decisively point at one of these two explanations.

\subsubsection*{SN 2017frh}
SN 2017frh was discovered by ASASSN and spectroscopically classified as a SN Ia. A flux excess was detected in the binned ZTF \ztfg~and \ztfr~bands at absolute magnitudes of $-$15.3 and $-$15.8, respectively. In the \ztfi-band, only upper limits were obtained. The detections were visible from the start of ZTF at a phase of 240 d after discovery and lasted for 180 d before disappearing behind the Sun. Nothing is detected after its return. The detections are found regardless of the baseline region that is used.

These detections are inconsistent with a normal SN Ia at these phases, which would have a significantly lower absolute magnitude ($M\approx-12$ mag) and also keep fading over time. If it were some type of sibling transient it would have to plateau at $-$15 to $-$16 mag for at least 180 d. Low luminosity SNe IIP can have a plateau in this magnitude range but typically last less long and become redder over time \citep{SN_II_colours} while the plateau in SN 2017frh becomes bluer over time. The luminous red nova AT 2021biy \citep{2021biy} does show a plateau of a similar duration but is much fainter than the late-time signal in SN 2017frh.

The late-time detections in SN 2017frh have a similar absolute magnitude and duration to the three late-time CSM interaction candidates that were presented in \citet{Terwel_2024_paper1}. However, the interaction in SN 2017frh starts significantly earlier ($\sim240$ d) than for those events. If we assume $F_R = 1.1$, a $M_\text{CSM}$ of 1.8 $M_\odot$ would be required to explain both the \ztfg- and \ztfr-bands assuming $T=10^5$~K. Raising the temperature to $T=10^7$~K gives $M_\text{CSM}\approx$ 4 $M_\odot$ for the \ztfg-band and $M_\text{CSM}\approx$ 4.5 $M_\odot$ for the \ztfr-band. SN 2017frh is shown in the left-hand panel of Fig.~\ref{M_CMS_fig}. If the $F_R$ is reduced to 1.01, the CSM mass is lowered to between 0.5 and 1.2 $M_\odot$. These mass estimates are in line with the suggestions of previous interacting SN events \citep{PTF11kx, Inserra_2016}. For these reasons, we classify SN 2017frh as a candidate SN with late-time CSM interaction.

\begin{figure*}
    \centering
    \includegraphics[width=\textwidth]{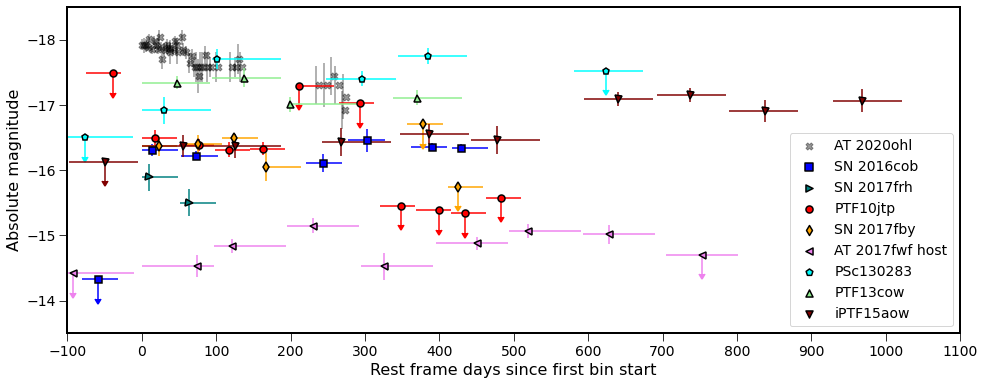}
    \caption{Binned \textit{r}-band observations of the objects whose late-time detections are consistent with a galaxy nucleus (see Table \ref{weirdos}) and are potentially transient events unrelated to the original SN. Upper limits are shown with downward arrows. The \textit{r}-band light curve of the slowly evolving ambiguous nuclear transient AT 2020ohl (ASASSN-20hx, \citealt{2020ohl_Hinkle}) is also shown for comparison. As the late-time signal in AT 2017fwf was found to be significantly offset from the SN location, we show the processed light curve at the host nucleus location instead.}
    \label{ANT_comp}
\end{figure*}

The late-time excess at the position of SN 2017frh is also on top of the host nucleus, which could mean that the late-time signal is instead related to the host. However, the duration of the detected signal is the shortest of those shown in Fig.~\ref{ANT_comp}, although there is an observing gap at the end of the detected signal so no strict limit can be placed. Late-time CSM interaction signatures have been found in SNe Ia at similar phases \citep{2015cp}. Therefore, given the relatively low CSM mass required and the time frame of the interaction, we prefer the CSM interpretation for this late-time signal.

\subsubsection*{SN 2017fby}
SN 2017fby was discovered by ASASSN and spectroscopically classified as a SN Ia. The ZTF binned \ztfr-band detections, at an absolute magnitude of $-$16.3 mag, are found when the sky location returns from behind the Sun at 1\,860 d after discovery and lasts for 200 d until it becomes unobservable again. The light curve stays roughly constant over this time frame. There are hints of a \ztfg-band excess as well but only one bin is above the $5\sigma$ threshold. After it comes back from behind the Sun nothing is detected anymore.

With a much stronger \ztfr\ band signal compared to the \textit{g} band, one could argue for this being a reddened previously unknown sibling transient whose rise was missed due to it occurring while the sky position was too close to the Sun to be observed. However, the long plateau in the light curve of at least 200 d rules this out, as Type IIP SNe have plateaus that are generally significantly shorter than this \citep{IIL_IIP, SN_II_V_band_lcs}.

CSM interaction with H-rich material could explain the strong \ztfr-band signal through strong \Halpha\ emission. If we assume the detections could come from a CSM shell with a fractional radius as low as $F_R \approx 1.005$, this gives a required CSM mass of $M_\text{CSM} \approx 5.5$ $M_\odot$ according to the calculation in Section \ref{CSM_calc}. This is around the upper end of the CSM masses that have been estimated for known SNe Ia-CSM, such as SN 1997cy \citep{Chugai_2004}, SN 2002ic \citep{Chugai_2004, Inserra_2016}, and SN 2012ca \citep{Inserra_2016}. Therefore, we note SN 2017fby as a SN with potentially late-time interaction with a thin, massive shell of CSM.

Checking the difference images with \textsc{snap} shows a clear bright spot at the SN location during the period of detections and no clear issues before or after it, confirming the excess to be real. At a distance of 0.73" (0.66 kpc at the host redshift) from the host nucleus, a TDE or other nuclear transient would likely appear on the same or adjacent pixel in the ZTF images. Like for SN 2016cob, this makes it difficult to say whether late-time CSM interaction or an ANT is the best explanation for the signal (Fig.~\ref{ANT_comp}).

\subsubsection*{PTF10jtp}
PTF10jtp was discovered by PTF and classified spectroscopically as a SN Ia. Detections were made in the ZTF binned \ztfr-band light curves starting at 2\,850 d after discovery. It had a $\sim 200$ day \ztfr-band plateau around $-$16.4 mag before the sky position became unobservable. When it came back from behind the Sun, there were $5\sigma$ upper limits that were about one magnitude deeper than the previous detections (not shown in Fig.~\ref{weirdo_plots}). 

The \ztfg\ band has upper limits at approximately $-$15.5 mag during this entire period, while the \ztfi\ band bins are not deep enough for a constraining upper limit to be placed. The baseline correction in \ztfr\ is significant and comparable between the later two baseline regions, but is somewhat smaller in the first region as it partially includes the period in which the excess was detected. This suggests that the excess was present in (some of) the reference images, and the late-time excess was present for longer than was found with the binning procedure. \textsc{snap} confirms this, as a slight excess can be seen in the difference images, followed by a small ghost at later times.

As this excess occurs over eight years after the SN, any SN Ia radioactive tail contribution at this magnitude can immediately be ruled out. Although there are only detections in one band, the limits placed on the colour together with the $>200$ day plateau limit the possibilities for a sibling, even with moderate values of extinction.

An unreasonably thin shell of CSM is needed to explain detections that are this late and still have an absolute magnitude of $M \approx -16.5$ mag while keeping the total mass of to be at most $M_\text{CSM} \approx 5$~$M_\odot$ in line with literature events. A more realistic shell thickness results in a CSM mass of at least 10 $M_\odot$. Therefore, we rule out CSM interaction as a likely explanation for this late-time flux excess. 

The sky location is 1.11\arcsec\ from the host nucleus, and \textsc{snap} also shows that the excess is slightly offset from the SN location in most frames and consistent with the host nucleus. The long duration of the excess, combined with the central location, suggests host activity is the cause, although it is not an AGN according to its WISE colours. Therefore, a nuclear transient is the most likely explanation for these late-time detections. Its light curve is shown compared to the ANT AT 2020ohl in Fig.~\ref{ANT_comp}, where it is seen to be plausibly consistent in duration and absolute magnitude.

\subsubsection*{AT 2017fwf}
This object was only detected by Gaia at early times with no spectrum obtained, preventing even a speculative classification based on the photometry. It is detected in ZTF binned \ztfr\ band data starting at 1\,750 d after discovery and lasts for 420 d. The light curve rises from $-$14 to nearly $-14.7$ mag before slowly fading again. Despite having upper limits around the same magnitude as \ztfr, nothing was detected in \ztfg. The \ztfi\ band limits are shallower and less constraining. \textsc{snap} shows that the excess is real, but located at the nucleus location 1.67\arcsec\ from the SN. Therefore, CSM interaction can be ruled out. The long timescale of the transient also makes a SN-like transient unlikely. 

Performing forced photometry at the host nucleus location reveals its properties much more clearly, being visible for nearly 700 days and reaching $-15.2$ mag at its brightest (see Fig.~\ref{ANT_comp}). This is longer by nearly 300 days and brighter by nearly 0.5 mag than the values given in Table \ref{weirdos} for the forced photometry at the SN position. The host galaxy is not an AGN according to its \textsc{wise} colours. The excess is still very faint for a TDE unless it is heavily reddened, although a reddened TDE would likely have been bright enough in the \ztfi\ band to still be visible. However, an ANT could be possible, though the signal is $\sim3$ mag fainter than AT 2020ohl but also visible for over twice as long. Faint nuclear variability could also be possible, and there is some scatter in the light curve, but with our 100 d bins for the light curve, small-scale variability can not be identified. Therefore, we can only conclude that some sort of nuclear variability is present. 

We note that a sibling transient is also detected by ZTF in this host galaxy (SN 2020ackb, SN IIP), though it is at a distance of 5.23\arcsec\ and has no effect on the forced photometry at the location of AT 2017fwf. The time at which the sibling was visible also does not correspond with the detected host variability.

\subsubsection*{PSc130283}
PSc130283 was detected by Pan-STARRS and spectroscopically classified as a SN Ia. At 4\,030 d past discovery, ZTF-detections are found in the \ztfr\ band at an absolute magnitude of $-$17 mag before brightening again to $-$17.8 mag in the next bin and staying there with variation $\leq 0.5$ mag for the remainder of the light curve (at least 520 d). The non-detections before the start of the excess were up to 1.5 mag lower, suggesting that the excess started suddenly rather than a gradual brightening. During the entire late-time excess, the \ztfg\ band stays with $5\sigma$ upper limits at $-$16 mag. At a redshift of $z=0.07622$, the \Halpha\ line is shifted into the overlap between the \ztfr- and \ztfi-bands. If the late-time signal is due to interaction with H-rich CSM, these detections could be a sign of strong \Halpha\ emission pushing the brightness in these bands.

Detections this late ($\gtrsim 3\,750$ rest frame days) and bright ($M \sim -18$ mag) cannot give a reasonable $M_\text{CSM}$ in Eq.~\ref{CSM_calc} unless a temperature around $10^{4.5}$~K and $F_R < 1.0005$ is assumed, giving an unreasonably thin shell and no extinction between the SN site and us. Therefore, we rule out CSM interaction as a likely cause of the late-time flux excess.

\textsc{snap} shows a clear excess that is consistent with the SN location, and no ghost or residuals before the excess starts or image defects that could explain these detections. However, the flux excess is only 0.44\arcsec\ from the host nucleus, although the host is not an AGN according to its WISE colours. The host has a history of small variability that causes sparse detections (including a detection that put its discovery date over 400 days before the SN explosion), which points to the ZTF detections in the binned data being host-related. Out of all excesses plotted in Fig.~\ref{ANT_comp}, PSc130283 is the brightest. It is very similar in brightness to AT 2020ohl though a bit longer in duration. These properties, combined with its relatively sharp rise suggest nuclear variability, such as an ANT, could explain these detections.

\subsubsection*{PTF13cow}
PTF13cow was discovered by PTF and classified as a SN Ia. The late-time detections in the ZTF \ztfg\ztfr\ztfi-bands begin 3\,230 d after the discovery at absolute magnitudes of $-$16.0 mag and $-$17.3 mag in the \ztfg- and \ztfr-bands, respectively. There are detections in the \ztfi-band, but the baseline is too small for these to be considered any further here. The \ztfr-band detections last for at least 610 d. The \ztfg-band is only detected at the start but is significantly fainter and disappears below the detection threshold earlier. This long timescale rules out SN-like transients as a cause for the excess.

\textsc{snap} shows that there is a clear excess in the images, and despite the host nucleus distance being only 0.56\arcsec\ from the SN location, the SN location and host nucleus are on different pixels. The excess has a preference to be at the host nucleus location, suggesting that it is the cause of the signal. However, the host nucleus is not an AGN according to its WISE colours. The duration of over 560 d and a peak absolute \ztfr-band magnitude of $-17.4$ mag of the identified late-time excess is similar to AT 2020ohl, meaning that an ANT could explain these detections.

\subsubsection*{iPTF15aow}
iPTF15aow was discovered by the intermediate PTF survey (iPTF) and classified as a SN Ia. The late-time detections begin in the \textit{r}-band at 2\,120 d after discovery at $-$16.4 mag. There are only upper limits in the \textit{g}- and \textit{i}- bands at this time, but at later times some detections in these bands are made. The detections last for at least 1\,050 d and slightly brighten over time to $-$17.2 mag in the last bins. The long-lived nature rules out SN-like transients at the same position as the original SN Ia. 

Even when assuming a scenario where the \ztfr\ band interaction stays at the level it was discovered at for the entire duration it was detected, $> 5$ M$_\odot$ of CSM is required in a very thin, low-temperature shell. However, the signal increases in strength over time, which would increase the required CSM mass. On top of that, no known SN Ia with CSM interaction has interaction as long as the late-time excess is found for or with the strength increasing over time, further disfavouring the CSM explanation.

The host nucleus is close by at 0.92\arcsec\ but from \textsc{snap} the transient does not look to significantly favour the SN or host nucleus location over the other. The host is not an AGN according to its WISE colours (W1 -- W2 = 0.077 mag, W2 -- W3 = 1.564 mag). 

The duration, shape, and close proximity to the host nucleus of the excess point to it being most likely host-related. It has a similar brightness to AT 2020ohl, but has a several times longer duration and is still ongoing. The fact that it continues to brighten is also very unusual. None of these signatures are consistent with what would be expected from a late-time CSM interaction.

\section{Discussion}
\label{discussion}
In this study, we found 98 cases with detections that cannot be explained as false positives due to observational, reduction, or software issues. These objects can be split into four groups: i) ongoing signatures of bright and/or nearby transients that were still detectable at the start of ZTF, ii) sibling transients at nearly the exact same location in the sky, iii) known variable sources such as AGN and variable stars, and iv) nine late-time flux excesses that required a more in-depth examination. Of these nine, we concluded that one was a sibling SN, five cases were nuclear transients close to the SN locations, two objects where it is unclear whether the ZTF signal is host or CSM-related, and one SN whose ZTF-detections are most consistent with late-time CSM interaction.

\subsection{Rarity of late-time signals from SNe}
We started our search for late-time signals based on the positions of 7\,718 transients discovered before ZTF. Table~\ref{recovered_obj_breakdown} shows the number of objects that had detections in ZTF, split over the six main types of SNe we distinguish between in our sample. In the sections below, we discuss the main conclusions for each of the likely classes (Tails, siblings, nuclear, CSM) of the late-time signals detected.

\begin{table}[]
   \caption{Number of recovered signals for each SN type in our sample.}
    \centering
    \resizebox{\columnwidth}{!}{
    \begin{tabular}{cccccccccc}
        \hline
        \hline
        Type & No. & \multicolumn{2}{c}{Tails} & \multicolumn{2}{c}{Siblings} & \multicolumn{2}{c}{Nuclear} &  \multicolumn{2}{c}{CSM}  \\
        && No. & \% & No. & \%& No. & \% \\
        \hline
        Ia & 4991 & 18 & 0.4 & 6 $^a$ & 0.1 & 7 & 0.1 & 3 & 0.1\\
        II & 1311 & 19 & 1.4 & 4 $^b$ & 0.3 &  0 & 0 & 0 & 0 \\
        Ib/c & 339 & 3 & 0.9 & 0 & 0 & 0 & 0 & 0 & 0 \\
        IIn & 279 & 8 & 2.9 & 0 & 0 & 0 & 0 & 0 & 0 \\
        I & 58 & 2 & 3.4 & 0 & 0 & 0 & 0 & 0 & 0 \\
       Ibn & 17 & 0 & 0 & 0 & 0 & 0 & 0 & 0 & 0 \\
        \hline
    \end{tabular}
    }
     \label{recovered_obj_breakdown}
    \tablefoot{
    As it is unclear whether SN 2016cob and SN2017fby should be counted as nuclear transients or due to CSM interaction, they have been counted in both.\\
    \tablefoottext{a}{This includes the unconfirmed but likely sibling at the position of SN 2017ige determined from our binned light curve analysis.\\}
    \tablefoottext{b}{This includes one sibling pair (SN 2009hz/AT 2018iml) where the ZTF transient was a likely CV based on its light curve.}
    }
\end{table}

\subsubsection{Long-lived transients}
We recovered 63 pre-ZTF transients that were still visible at the start of ZTF. In all these cases, the transient is in the reference images as well, resulting in a ghost in the difference images. The only way to recover these transients is by performing a proper baseline correction to correct for the SN being in the references. 

50 transients in this group are spectroscopically classified SNe with (slowly) declining tails that are bright enough to be detected by ZTF for hundreds of days. Several of these have been studied in more detail (see Table~\ref{tail_objects} for references to papers discussing these). The other 13 objects are SN candidates: transients that are likely SNe based on their photometry but were never spectroscopically classified. One of these is AT~2017gpv, which is notable for having a persistent signal up to more than 1400 days after it was first detected and its similarity to iPTF14hls.

\subsubsection{Siblings}
Within the uncertainties, the rate of siblings we found is the same for the different groups of objects, with the SN Ia and Type II SN classes having rates of sibling discovery of 0.1 -- 0.2 \%. The percentage is similar, given the large uncertainties associated with low numbers of events, to the percentage of siblings found in \citet{Terwel_2024_paper1} over a smaller time frame. These rates can be taken as lower limits because of observing gaps, which would result in transients being missed. In total we found 10 siblings, which gives a lower limit on the rate of siblings of $0.13\pm0.04\%$. SN Ia sibling pairs along the same line of sight may be useful for constraining the origin of reddening in SN Ia light curve fitting for cosmology \citep{DR2_siblings}, but a larger sample would be required for meaningful constraints.

\subsubsection{Serendipitous nuclear transient detections}
\label{sec:discussion:nuclear}
We identified five late-time signals that were most likely due to host activity that was (partially) picked up in the forced photometry at the SN location. This was determined due to a combination of the shape and duration of the late-time light curve, as well as the inferred $M_\text{CSM}$ from the absolute magnitude and delay between the main SN and late-time detections being unrealistically high in the best-case scenario. We also identify signals at the positions of two events (SN 2016cob and SN 2017fby), where the inferred cause could be CSM or nuclear activity, bringing the total to seven potential nuclear events (see Table \ref{recovered_obj_breakdown}).

In Fig.~\ref{ANT_comp}, we show the late-time excess light curves for which nuclear transients can not be ruled, along with the late-time light curve of SN 2017frh for completeness, although we prefer a CSM interpretation for it. The light curves are shown as absolute magnitude in the \ztfr\ band against the time since the first binned data where a flux excess is identified. We also show the ANT AT 2020ohl \citep{2020ohl_Hinkle} for comparison. AT 2020ohl is slightly brighter than most of these transient events, apart from PSc130283. Our events are also longer lasting, ranging from $\sim200$ days up to $\sim1\,100$ days. This could mean that there is a previously unknown population of lower luminosity ANTs that require deeper (binned) observations to be detected.

\citet{wiseman_ztfants} showed that their sample of ANTs all have changes in their WISE \textit{W1}- and \textit{W2}-band observations, with a slight delay compared to the optical. We investigated the WISE light curve for our potential nuclear transients, but unfortunately, these light curves use the total measured flux from the source. As the hosts of our nuclear transients have an apparent magnitude of around 15 mag in these bands and the optical variability we have identified is $\geq5$ mag fainter than that, any MIR variability is likely to be heavily suppressed into the noise of the WISE light curves. \citet{WISE_diff_im} have recently built an image-subtraction pipeline for WISE / NEOWISE data, which could be used to solve this problem.

CL-AGN show variability on similar timescales as those that we have found here \citep{CLAGN}. However, CL-AGN are a subset of AGN and the galaxies hosting our transients are not AGN according to the WISE criterion from \citet{WISE_crit}. Our objects are also much lower luminosity than seen for CL-AGN, which also makes this classification unlikely.

\subsubsection{CSM interaction}
In \citet{Terwel_2024_paper1}, we estimated an intrinsic rate, through simulations of the detection efficiency for three out of 3\,628 SNe Ia with potential late-time CSM signals, of $<$0.5 per cent of normal SNe Ia displaying late-time ($>$ 100 d after the peak) CSM signatures. \cite{GALEX_Late_CSM} estimated a rate of late-time interaction in $<$5 \% of SNe Ia. In this work, we identified three SNe Ia whose signals in ZTF cannot be ruled out as being due to late-time CSM interaction, giving a raw rate of late-onset CSM interaction of $\leq$0.1 \% for the 4\,991 SNe Ia in this sample (see Table \ref{recovered_obj_breakdown}). No potential late-time signature from CSM interaction was identified in any other SN type.

SN 2017frh is our best candidate for late-onset CSM interaction. It has the earliest period of late-time detections ($\sim 250$ days after the first detection) and a similar estimate for $M_\text{CSM}$ as \citet{2015cp} found for SN 2015cp. If we assume a nova-like shell at $T=10^7$~K, we get a reasonable CSM mass of $M_\text{CSM} \approx 4$ $M_\odot$ (although see the discussion of the limitations of our CSM estimates in Section \ref{sec:discussion:limitations}). Our two other CSM candidates, SN 2017fby and SN 2016cob, have late-time signals five years after the SN was first detected. When we assume these objects to have shells of similar thickness as the shell in SN 2017frh, a CSM mass estimate of $\sim 5$ $M_\odot$ can be found for these objects as well. These estimates are lower limits, as no host extinction is assumed for objects in regions of galaxies where significant amounts of extinction can be expected.

We have not performed detailed rate calculations for the events discovered in this study. However, there are $\sim600$ objects in our sample that were first detected less than 300 days before the start of ZTF, and we have one candidate with a late-time signal detected around 300 days after discovery. This gives a raw rate of 0.2 \% when not taking into account any bias. Two additional candidates have their detections in the fifth year after the explosion. About half of our sample exploded less than five years before the start of ZTF, and have their fifth year observed the survey. Two of these are candidates for having late-time CSM interaction at these phases, giving a rate of 0.05\% when not taking any bias into account. Our rough estimates of the rates suggest similar values to those of \citet{Terwel_2024_paper1}, within the uncertainty of the small numbers of events detected.

The three events in \citet{Terwel_2024_paper1} that were identified as consistent with CSM interaction were also close to their host nuclei. As discussed in \citet{Terwel_2024_paper1}, this could suggest a preference for CSM interaction to occur in SNe Ia in these environments or an alternative explanation is that they are caused by nuclear activity. However, as discussed in Section \ref{sec:discussion:nuclear} and shown in Fig.~\ref{ANT_comp}, these late-time signals are similar but not entirely consistent with nuclear activity/transients.

\subsection{Limitations}
\label{sec:discussion:limitations}
We found a large group of SNe whose tail was detected by ZTF but never properly picked up by the survey -- all of them required more thorough baseline corrections as the transient was present in the reference images. Several of these were exceptionally long-lived, with detections in the binned data being recovered over a year after the start of ZTF. While some of these can be explained by CSM interaction slowing the decline of the light curve, the gaps between the pre-ZTF and ZTF detections can make it difficult to say whether the late-time signal is due to new interaction or if the transient has been steadily fading between the observations. Some objects were already known to be interacting before the start of ZTF but we still grouped them with the tail detections as their interaction did not start at late times. We also found some periodic variable sources that have this behaviour for at least the length of the survey, making it impossible to generate proper reference images or apply a good baseline correction. Both groups can leave behind a ghost, which is an imprint at the transient location when the object is fainter than it was in the references. These events can be recovered from the data, particularly if different time frames for estimating the baseline correction are used, but they are often hidden if a baseline correction is not applied. 

As in \citet{Terwel_2024_paper1}, the magnitude limit of the binning technique is set by the magnitude limit of the references, as their uncertainty starts to dominate when more epochs of observation are binned together. One way to improve the sensitivity, though computationally expensive, is to find a region where we are certain there is no excess at the SN location and use it to generate custom references to use for difference imaging. This would immediately lessen the need for a baseline correction significantly, as an adequately chosen reference region already has this built-in automatically.

Any search for late-time signals from SNe will also be dependent on the constraints placed on the sample. We required a first detection between 2008 and 2018 and a classification or candidate SN status. The signals must be bright enough and last long enough to be picked up after binning the ZTF data, setting constraints on the minimal strength and duration of these signals. Our sample selection does not allow us to give an intrinsic rate of late-time CSM interaction, but it does show the difficulty of finding these events.

Our sample also contained several false positives from objects that had spurious detections within the initial sample definition, but a SN (which was classified) exploded years later and was observed by ZTF. Siblings, whether they are true SN siblings or false due to spurious detections or projection effects, can be confused if they occur at (nearly) the same position in the sky. One way to remove false siblings such as AT 2017keg and SN 2013ld is to require a spectroscopic classification in the same time period as the first detection, though this would also remove objects like AT 2017gcd as in that case only the second sibling was classified, and candidate objects like AT 2017fwf and AT 2017gpv that were never spectroscopically classified at all.

We attempted to estimate the CSM mass required to produce late-time detections as bright as we observed them, assuming Bremsstrahlung to be the main mechanism producing light. In most cases when assuming a thin nova-like shell the required CSM mass is above $\sim 10$ $M_\odot$, with the only exception being SN 2017frh due to its relatively short delay time between the SN explosion and late-time detection. For SN 2016cob and SN 2017fby, the only way to get realistic CSM mass estimates is by assuming a colder CSM and a lower $F_R$, which could be explained by the shell size that is needed for interaction to start over five years after the SN (see the end of Sect.~\ref{CSM_calc}). \citet{2015cp} argue that the simplified assumption of Bremsstrahlung being the main mechanism is largely responsible for the high CSM mass estimates. They find an upper limit of $M_\text{CSM} \sim 7$ $M_\odot$ for SN 2015cp with this method, while \citet{2015cp_radio} find $M_\text{CSM} \lesssim 0.5$ $M_\odot$ using radio non-detections. They argue that line emission from elements such as \ion{H}, \ion{Mg}, and \ion{Ca}~are the main emission contributors instead. However, as line emission requires much more assumptions on the state of the emitting gas, it is much more complicated to derive a good estimation for $M_\text{CSM}$ for it. 

Even if we assume that Eq.~\ref{M_CSM_eq} gives an upper limit and the better estimates from modelling of radio observations to be $\sim14$ times smaller (based on the two estimates for SN 2015cp), the CSM interaction scenario for late-time signatures at 2\,000 -- 4\,000 days after the initial SN discovery is still unlikely for five of our events that are also potentially due to nuclear activity. We have also not considered the impact of extinction in the host galaxy, which is likely non-negligible given the nuclear location of the transients considered. The CSM mass can be reduced by assuming a smaller $F_R$, but the effect is limited unless we consider unphysically thin shells. However, thin shells are not ruled out by a several hundred-day long period of late-time detections, as the time delay due to the large radius of the shell can smear even a 1~s interaction signal out over several months. This shows that it is very difficult to properly constrain the CSM mass from photometry alone without knowing the mechanism powering it and more detailed theoretical studies.

\section{Conclusions}
\label{conclusions}
In this work we have presented a search for late-time signals observed in ZTF from transients that were first detected up to ten years before the start of the survey. By binning the ZTF data we were able to go nearly one magnitude below the noise limit of unbinned data points. Our sample consists of 7\,718 unique objects that are in the ZTF footprint and could be searched for late-time signatures. By careful inspection of the ZTF images and comparison to known classes of transients, we determined the most likely source of the late-time signatures we identified at the positions of 98 transients discovered prior to the start of ZTF. Our main conclusions are:

\begin{enumerate}
    \item We identified several very long-lived transients that began pre-ZTF and continued for hundreds to thousands of days. These include the SLSN IIn, SN 2015da, with data extending to eight years after discovery, and the spectroscopically unclassified AT 2017gpv, whose bumpy long-lived light curve (detected up to nearly four years after the discovery) is reminiscent of iPTF14hls \citep{iPTF14hls_Iair, Sollerman_2019_iptf14hls} and SN 2020faa \citep{Yang_2021_20faa, 2020faa_hidden_shocks}. 
    \item We found ten confirmed pairs of sibling transients where the pre-ZTF and ZTF siblings are at nearly the same sky position. One additional unconfirmed but likely sibling transient was identified when applying our binning technique at the positions of SN 2017ige.
    In most cases, one of the siblings was never spectroscopically classified. We also found two cases where the ZTF transient was mistakenly reported as a pre-ZTF transient due to spurious detections from the nearby host.
    \item We found six flux excesses at the position of pre-ZTF events classified as Type II SNe (three Type II, one IIn and two SLSN-II) consistent with AGN activity in known AGN host galaxies. We speculate that some of these were likely misclassified at the time of the pre-ZTF detection.
    \item We found five flux excesses that were consistent with activity of the host nucleus close to the location of the original transient. The brightness, light curve shape, and duration are broadly consistent with (ambiguous) nuclear transients, showing that there may be a previously unknown population of faint nuclear transients that requires deeper (binned) observations to be discovered.
    \item We found three flux excesses that were consistent with late-time interaction with a $4 - 5\ M_\odot$ CSM shell, assuming Bremsstrahlung as the main emission mechanism. In two of these, the signal is $\sim5$ years after the SN and could also be explained by a nuclear transient unrelated to the SN. Our best candidate is SN 2017frh with a signal $\sim250$ days after the SN, consistent with a $4\ M_\odot$ nova-like CSM shell at $10^7$ K. However, a nuclear transient can not be completely ruled out.
 \end{enumerate}

The Vera C. Rubin Observatory's Legacy Survey of Space and Time \cite[LSST;][]{LSST} will observe the transient universe several magnitudes deeper than ZTF. This is crucial for finding faint interaction signals like those we have searched for in this paper, as we have shown these to be rare. Increasing the volume that can be probed is the best way to find new events and follow them up. Current samples of objects will be observed at late times by LSST, making it possible to do a late-time signal search in real time. By using a well-defined and complete sample such as the ZTF SN Ia DR2 \citep[][, Smith et al.~in prep.]{DR2_Overview}, Bright Transient Survey \cite[BTS;][]{BTS_I, BTS_II}, or a similar sample for ZTF SNe first detected in 2021+ as the basis for a similar search of late-time signals observed with LSST, it will also be possible to properly estimate their rates. Not only can these methods be used to search for late-time CSM interaction, but as we have shown, they are also suitable for finding other weak signals, such as faint nuclear transients.

\begin{acknowledgements}
We thank Peter Clark, Morgan Fraser, and Mariusz Gromadzki for helpful discussions.

JHT and KM acknowledge support from EU H2020 ERC grant no. 758638.
PW acknowledges support from the Science and Technology Facilities Council (STFC) grants ST/R000506/1 and ST/Y001850/1.

Based on observations obtained with the Samuel Oschin Telescope 48-inch and the 60-inch Telescope at the Palomar Observatory as part of the Zwicky Transient Facility project. ZTF is supported by the National Science Foundation under Grants No. AST-1440341 and AST-2034437 and a collaboration including current partners Caltech, IPAC, the Oskar Klein Center at Stockholm University, the University of Maryland, University of California, Berkeley, the University of Wisconsin at Milwaukee, University of Warwick, Ruhr University, Cornell University, Northwestern University and Drexel University. Operations are conducted by COO, IPAC, and UW.

The Pan-STARRS1 Surveys (PS1) and the PS1 public science archive have been made possible through contributions by the Institute for Astronomy, the University of Hawaii, the Pan-STARRS Project Office, the Max-Planck Society and its participating institutes, the Max Planck Institute for Astronomy, Heidelberg and the Max Planck Institute for Extraterrestrial Physics, Garching, The Johns Hopkins University, Durham University, the University of Edinburgh, the Queen's University Belfast, the Harvard-Smithsonian Center for Astrophysics, the Las Cumbres Observatory Global Telescope Network Incorporated, the National Central University of Taiwan, the Space Telescope Science Institute, the National Aeronautics and Space Administration under Grant No. NNX08AR22G issued through the Planetary Science Division of the NASA Science Mission Directorate, the National Science Foundation Grant No. AST-1238877, the University of Maryland, Eotvos Lorand University (ELTE), the Los Alamos National Laboratory, and the Gordon and Betty Moore Foundation.

This work has made use of data from the European Space Agency (ESA) mission
{\it Gaia} (\url{https://www.cosmos.esa.int/gaia}), processed by the {\it Gaia}
Data Processing and Analysis Consortium (DPAC,
\url{https://www.cosmos.esa.int/web/gaia/dpac/consortium}). Funding for the DPAC
has been provided by national institutions, in particular the institutions
participating in the {\it Gaia} Multilateral Agreement.

The intermediate Palomar Transient Factory project is a scientific collaboration among the California Institute of Technology, Los Alamos National Laboratory, the University of Wisconsin (Milwaukee), the Oskar Klein Centre, the Weizmann Institute of Science, the TANGO Program of the University System of Taiwan, and the Kavli Institute for the Physics and Mathematics of the Universe.

ASAS-SN is funded by Gordon and Betty Moore Foundation grants
GBMF5490 and GBMF10501 and the Alfred P. Sloan Foundation
grant G-2021-14192.

The Gordon and Betty Moore Foundation, through both the Data-Driven Investigator Program and a dedicated grant, provided critical funding for SkyPortal.

The ztfquery code was funded by the European Research Council (ERC) under the European Union's Horizon 2020 research and innovation programme (grant agreement n°759194 - USNAC, PI: Rigault).
\end{acknowledgements}

\section*{Data Availability}
The full sample of objects used in this paper can be retrieved from \url{https://github.com/JTerwel/pre-ZTF_transients}, and the original object data can be retrieved from the Open Supernova Catalog\footnote{\url{https://github.com/astrocatalogs/supernovae}} and WISeREP\footnote{\url{https://www.wiserep.org}}. The ZTF light curves were generated using \textsc{fpbot}\footnote{\url{https://github.com/simeonreusch/fpbot}}. The binning program can be found at \url{https://github.com/JTerwel/late-time_lc_binner}, and \textsc{snap} can be found at \url{https://github.com/JTerwel/SuperNova_Animation_Program}.

\bibliographystyle{aa}
\bibliography{aanda}

\begin{appendix}
\onecolumn

\section{Tables}
\begin{longtable}{cccccc}
\caption{List of the 63 pre-ZTF objects whose tail is still visible in ZTF using our binning method after a baseline correction.}
 \label{tail_objects}
 \endfirsthead
 \caption{continued}\\
 \hline
 \hline
 Name &  Type & $z$ & Discovery date & visible until (mjd) & Object Papers \\
 \hline
 \endhead
 \hline
 \endfoot
 \hline
 \endlastfoot
 \hline
 \hline
 Name &  Type & $z$ & Discovery date & visible until (mjd) & Object Papers \\
 \hline
 AT 2017fwg & Candidate & 0.113627 & 01-08-2017 & 58325 & \\
 AT 2017gpv & Candidate & 0.025563 & 04-09-2017 & 59480 & \\
 AT 2017gpy & Candidate & 0.025187 & 04-09-2017 & 58250 & \\
 AT 2017hat & Candidate & 0.020454 & 01-10-2017 & 58275 & \\
 AT 2017igk & Candidate & 0.040847 & 16-11-2017 & 58250 & \\
 AT 2017ihh & Candidate & 0.027521 & 03-10-2017 & 58300 & \\
 AT 2017iho & Candidate & 0.053421 & 17-11-2017 & 58250 & \\
 AT 2017ims & Candidate & 0.030094 & 15-11-2017 & 58300 & \\
 AT 2017iru & Candidate & 0.018156 & 29-11-2017 & 58375 & \\
 ATLAS17nbe & Candidate & 0.013476 & 05-11-2017 & 58400 & \\
 ATLAS17oai & Candidate & 0.099205 & 23-12-2017 & 58300 & \\
 ATLAS18eas & Candidate & 0.033103 & 30-12-2017 & 58325 & \\
 ATLAS18mmr & Candidate & 0.077507 & 15-12-2017 & 58325 & \\
 DES16X2bkr & SN II & 0.159 & 21-09-2016 & 58400 & \\
 SN 2015da & SLSN IIn & 0.007222 & 09-01-2015 & 59475 & \citet{2015da_2020, 2015da_2024}\\
 \multirow{2}{*}{SN 2016bkv} & \multirow{2}{*}{SN II} & \multirow{2}{*}{0.002} & \multirow{2}{*}{21-03-2016} & \multirow{2}{*}{58475} & \citet{2016bkv_Hosseinzadeh};\\
 &&&&&\citet{2016bkv_Nakaoka, 2016bkv_Deckers}\\
 SN 2016cyi & SN IIn & 0.044 & 25-06-2016 & 58475 & \\
 SN 2016ieq & SN IIn & 0.066 & 14-11-2016 & 58400 & \\
 SN 2017aym & SN IIP & 0.005928 & 13-01-2017 & 58700 & \\
 SN 2017dio & SN Ic & 0.037 & 26-04-2017 & 58300 & \citet{2017dio, 2017dio_Shi}\\
 SN 2017dpu & SN II & 0.018 & 30-04-2017 & 58250 & \\
 SN 2017eby & SN Ia-CSM & 0.081 & 01-04-2017 & 58350 & \\
 \multirow{6}{*}{SN 2017egm} & \multirow{6}{*}{SLSN I} & \multirow{6}{*}{0.030721} & \multirow{6}{*}{23-05-2017} & \multirow{6}{*}{58300} & \citet{2017egm_Chen, 2017egm_Nicholl};\\
 &&&&&\citet{2017egm_Wheeler, 2017egm_Bose};\\
 &&&&&\citet{2017egm_Izzo, 2017egm_Maund};\\
 &&&&&\citet{2017egm_Hatsukade, 2017egm_Saito};\\
 &&&&&\citet{2017egm_Tsvetkov, 2017egm_Lin};\\
 &&&&&\citet{2017egm, 2017egm_Shang}\\
 SN 2017emq & SN Ia & 0.005247 & 03-06-2017 & 58250 & \\
 SN 2017erp & SN Ia & 0.006174 & 13-06-2017 & 58375 & \citet{2017erp}\\
 SN 2017err & SLSN IIn & 0.107 & 12-06-2017 & 58275 & \\
 SN 2017faa & SN II & 0.01845 & 27-06-2017 & 58300 & \\
 SN 2017fgc & SN Ia & 0.007722 & 11-07-2017 & 58400 & \citet{2017fgc_Burgaz, 2017fgc_Zeng}\\
 SN 2017fvr & SN IIP & 0.012539 & 01-08-2017 & 58450 & \\
 SN 2017gas & SN IIn & 0.011 & 10-08-2017 & 58850 & \citet{2017gas_Ransome, 2017gas_Bilinski}\\
 SN 2017ghw & SN IIn & 0.076 & 25-08-2017 & 58450 & \\
 SN 2017glx & SN Ia-91T & 0.011294 & 03-09-2017 & 58400 & \\
 \multirow{2}{*}{SN 2017gmr} & \multirow{2}{*}{SN IIP} & \multirow{2}{*}{0.005037} & \multirow{2}{*}{04-09-2017} & \multirow{2}{*}{58475} & \citet{2017gmr, 2017gmr_Nagao};\\
 &&&&&\citet{2017gmr_Utrobin}\\
 SN 2017gvb & SN IIn & 0.030344 & 18-09-2017 & 59000 & \\
 SN 2017gww & SN II & 0.01748 & 26-09-2017 & 58375 & \\
 SN 2017gxq & SN Ia & 0.008406 & 17-09-2017 & 58450 & \\
 SN 2017hbg & SN II & 0.016 & 25-09-2017 & 58375 & \\
 SN 2017hca & SN II & 0.013403 & 28-09-2017 & 58450 & \\
 \multirow{3}{*}{SN 2017hcc} & \multirow{3}{*}{SN IIn} & \multirow{3}{*}{0.0173} & \multirow{3}{*}{02-10-2017} & \multirow{3}{*}{58500} & \citet{2017hcc_Prieto, 2017hcc_Kumar};\\
 &&&&&\citet{2017hcc_Smith, 2017hcc_Chandra};\\
 &&&&&\citet{2017hcc_Moran, 2017hcc_Mauerhan}\\
 SN 2017hfv & SN Ia & 0.028199 & 10-10-2017 & 58200 & \\
 SN 2017hix & SN Ic & 0.012 & 13-10-2017 & 58375 & \\
 SN 2017hlt & SN Ia & 0.027 & 10-10-2017 & 58275 & \\
 SN 2017hmi & SN Ia & 0.0398 & 18-10-2017 & 58250 & \\
 SN 2017hpa & SN Ia & 0.015654 & 25-10-2017 & 58400 & \citet{2017hpa_Dutta, 2017hpa_Zeng}\\
 SN 2017hqj & SN IIP & 0.009 & 27-10-2017 & 58300 & \\
 SN 2017hro & SN II & 0.015 & 28-10-2017 & 58800 & \\
 SN 2017igf & SN Ia & 0.005624 & 12-11-2017 & 58350 & \\
 SN 2017ijr & SN Ia & 0.04 & 20-11-2017 & 58300 & \\
 SN 2017ijx & SN Ia & 0.027729 & 18-11-2017 & 58300 & \\
 SN 2017ivh & SN II & 0.008 & 05-12-2017 & 58300 & \\
 SN 2017ivu & SN IIP & 0.006528 & 11-12-2017 & 58400 &\\
 SN 2017ivv & SN II & 0.022 & 12-12-2017 & 58450 & \citet{2017ivv_gutierrez}\\
 SN 2017ixg & SN Ia & 0.0277 & 14-12-2017 & 58400 & \\
 SN 2017ixv & SN Ic-BL & 0.007302 & 17-12-2017 & 58350 & \\
 SN 2017ixx & SN II & 0.041 & 17-12-2017 & 58300 & \\
 SN 2017ixz & SN IIb & 0.024 & 14-12-2017 & 58250 & \\
 SN 2017iyd & SN IIb & 0.0285 & 13-12-2017 & 58275 & \\
 SN 2017jav & SN Ia & 0.01517 & 19-12-2017 & 58275 & \\
 SN 2017jbj & SN II & 0.013492 & 20-12-2017 & 58400 & \\
 SN 2017jeh & SN Ia & 0.020961 & 26-12-2017 & 58275 & \\
 SN 2018L & SN Ia & 0.02582 & 25-12-2017 & 58325 & \\
 SN 2018bq & SN Ia & 0.025628 & 30-12-2017 & 58275 & \\
 SN 2018fd & SLSN I & 0.263 & 11-10-2017 & 58450 & \\
\end{longtable}
\tablefoot{For transients that were not spectroscopically classified the quoted redshift is that of the host galaxy.}

\newpage
\section{Recovered non-SN sources}
\begin{figure}[h!]
    \centering
    \includegraphics[width=0.9\textwidth]{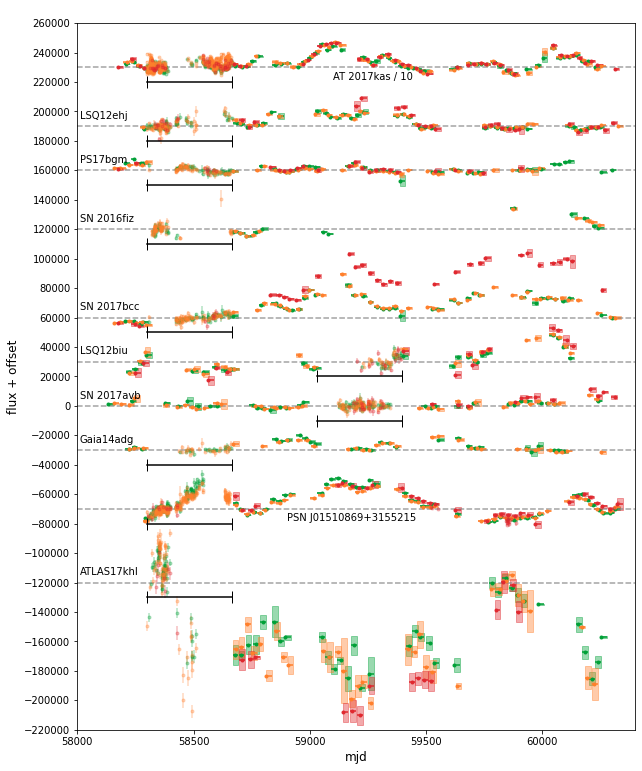}
    \caption{Binned flux of the recovered AGN. The flux is calibrated to a zeropoint at mag 30. The dashed lines show the baseline value for each object. Since the observations used to determine the baseline cannot be binned, we show the unbinned data in the baseline region of each object. This region is marked by the black lines. Even though not all objects are properly sampled over the entire lifetime of ZTF, they all clearly show variability over long timescales. As these objects are always varying, it is impossible to do a baseline correction without the transient present. This causes some light curves to go below their baseline. Note that the values for AT 2017kas have been divided by 10 as the variability is so large.}
    \label{non-transients_AGN}
\end{figure}

\begin{figure}[h!]
    \centering
    \includegraphics[width=0.9\textwidth]{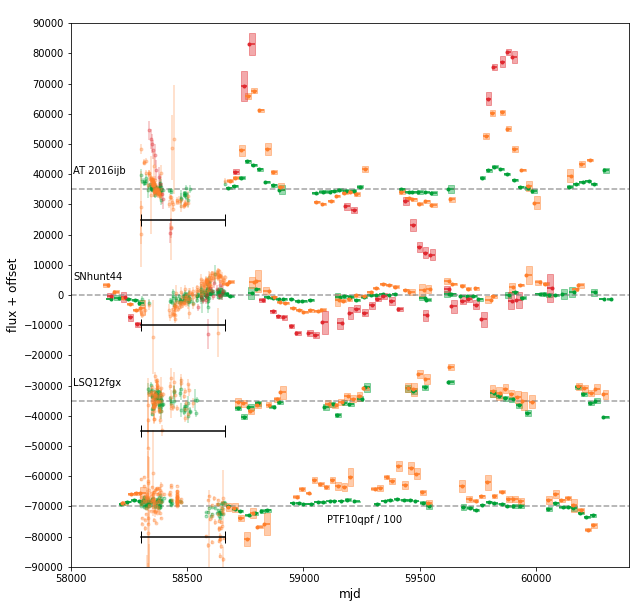}
    \caption{Binned flux of the recovered variable stars. The flux is calibrated to a zeropoint at mag 30. The dashed lines show the baseline for each object. Since the observations used to determine the baseline cannot be binned, we show the unbinned data in the baseline region of each object. This region is marked by the black lines. All three objects vary around their central values, but the amplitude varies widely between the ZTF bands. Note that the values for PTF10qpf have been divided by 100 as the variability is so large.}
    \label{non-transients_varstar}
\end{figure}

\end{appendix}
\end{document}